\documentclass[sigconf,authorversion,nonacm,pdfa]{acmart}

\usepackage{euscript}
\usepackage{mathtools}
\usepackage{listings}
\usepackage{cleveref}
\usepackage{algorithm}
\usepackage{algpseudocode} 
\usepackage{makecell}
\usepackage{enumerate}
\usepackage{enumitem}
\usepackage[skip=0.2pt]{subcaption}

\usepackage{eucal}

\newtheorem{example}{Example}[section]

\algrenewcommand\algorithmiccomment[1]{\hspace{2mm}\textcolor{blue}{$\triangleright$ #1}}
\algrenewcommand\algorithmicrequire{\textbf{Input:}}
\algrenewcommand\algorithmicensure{\textbf{Output:}}

\definecolor{dkgreen}{rgb}{0,0.6,0}
\definecolor{gray}{rgb}{0.5,0.5,0.5}
\definecolor{mauve}{rgb}{0.58,0,0.82}

\newcommand{\common}[1]{#1}
\newcommand{\reva}[1]{#1}
\newcommand{\revb}[1]{#1}
\newcommand{\revc}[1]{#1} 

\lstdefinestyle{sqlstyle}{
  language=SQL,
  basicstyle={\footnotesize\ttfamily},
  aboveskip=0mm,
  belowskip=1mm,
  breakatwhitespace=true,
  breaklines=true,
  classoffset=0,
  columns=flexible,
  commentstyle=\color{dkgreen},
  framexleftmargin=0.25em,
  frameshape={}{yy}{}{}, 
  keywordstyle=\color{blue},
  numbers=none, 
  numberstyle=\tiny\color{gray},
  showstringspaces=false,
  stringstyle=\color{mauve},
  tabsize=2,
  xleftmargin =1em,
  escapeinside=``
}

\newcounter{sqlcounter}

\crefname{listing}{Listing}{Listings}

\lstnewenvironment{sqllisting}[3][]{%
  \setcounter{lstlisting}{\value{sqlcounter}}%
  \lstset{style=sqlstyle,frame=tb,mathescape=true,caption={#2},label={#3},#1}%
}{%
  \stepcounter{sqlcounter}%
}

\newcommand\vldbdoi{10.14778/3819518.3819567}
\newcommand\vldbpages{2508-2521}
\newcommand\vldbvolume{19}
\newcommand\vldbissue{9}
\newcommand\vldbyear{2026}
\newcommand\vldbauthors{\authors}
\newcommand\vldbtitle{\shorttitle} 
\newcommand\vldbavailabilityurl{https://github.com/louisja1/bacon}
\newcommand\vldbpagestyle{empty}

\let\oldsubparagraph\subparagraph
\renewcommand{\subparagraph}[1]{\smallskip\oldsubparagraph{\mbox{}\hspace*{1em}\itshape(#1)}\mbox{}}

\newcommand{\narrow}[1]{\ensuremath{\text{\scalebox{0.9}[1.0]{\textsf{\upshape#1}}}}}
\newcommand{\myvec}[1]{\ensuremath{\smash{\vec{#1}}}}
\newcommand{\concat}{\ensuremath{\mathbin{\smallsmile}}}

\newcommand{\sql}[1]{\narrow{#1}}

\newcommand{\selection}{\sigma}
\newcommand{\join}{\Join}

\newcommand{\Tables}[1]{\ensuremath{\narrow{Tables}_{#1}}}
\newcommand{\Attrs}[1]{\ensuremath{\narrow{Attrs}{[#1]}}}
\newcommand{\JPreds}[1]{\ensuremath{\narrow{Preds}^{\join}_{#1}}}
\newcommand{\JAttrs}[1]{\ensuremath{\narrow{Attrs}^{\join}_{#1}}}
\newcommand{\SAttrs}[1]{\ensuremath{\narrow{Attrs}^{\selection}_{#1}}}
\newcommand{\SPreds}[1]{\ensuremath{\narrow{Preds}^{\selection}_{#1}}}

\newcommand{\database}{\ensuremath{\mathcal{D}}}
\newcommand{\tab}{\ensuremath{R}}
\newcommand{\tabs}{\ensuremath{\mathcal{R}}}

\newcommand{\attr}{\ensuremath{A}}
\newcommand{\attrs}{\ensuremath{\mathcal{A}}}
\newcommand{\queryset}{\ensuremath{\pmb{\mathcal{Q}}}}

\newcommand{\query}{\ensuremath{Q}}

\newcommand{\equivalentclass}{\ensuremath{\mathcal{E}}}

\newcommand{\batchparam}{\ensuremath{\beta}}

\newcommand{\CountMap}{\ensuremath{\mathfrak{M}}}
\newcommand{\QuantScale}{\ensuremath{\mathfrak{b}}}
\newcommand{\QuantScales}{\ensuremath{\mathfrak{B}}}

\newcommand{\sequentialproc}{\textsc{IndProc}}
\newcommand{\postfilter}{\textsc{PostFilt}}
\newcommand{\ours}{\textsc{BaCon}}

\newcommand{\hybrid}{\textsc{Hybrid}}

\newcommand{\ourrecursive}{\ours\textsc{Recurse}}
\newcommand{\optourrecursive}{\batchparam\textsc{-}\ourrecursive}

\newcommand{\jp}{\ensuremath{\mathfrak{J}}}
\newcommand{\njps}{\ensuremath{q}}

\newcommand{\nxt}{\ensuremath{\text{next}}}

\newcommand{\ProcessTable}{\textsc{ProcessTable}}
\newcommand{\OptProcessTable}{\batchparam\textsc{-}\ProcessTable}
\newcommand{\ConstructQuantScales}{\textsc{QuantScales}}
\newcommand{\ComputeCounts}{\textsc{ComputeCounts}}
\newcommand{\treeroot}{\narrow{root}}
\newcommand{\subtree}{\narrow{subtree}}
\newcommand{\children}{\narrow{children}}
\newcommand{\parent}{\narrow{parent}}

\newcommand{\Jin}{\ensuremath{J^\narrow{in}}}
\newcommand{\Jout}{\ensuremath{J^\narrow{out}}}

\definecolor{lstpurple}{rgb}{0.5,0,0.5}
\definecolor{lstred}{rgb}{1,0,0}
\definecolor{lstreddark}{rgb}{0.7,0,0}
\definecolor{lstredl}{rgb}{0.64,0.08,0.08}
\definecolor{lstmildblue}{rgb}{0.66,0.72,0.78}
\definecolor{lstblue}{rgb}{0,0,1}
\definecolor{lstmildgreen}{rgb}{0.42,0.53,0.39}
\definecolor{lstgreen}{rgb}{0,0.5,0}
\definecolor{lstorangedark}{rgb}{0.6,0.3,0}	
\definecolor{lstorange}{rgb}{0.75,0.52,0.005}
\definecolor{lstorangelight}{rgb}{0.89,0.81,0.67}
\definecolor{lstbeige}{rgb}{0.90,0.86,0.45}

\DeclareFontShape{OT1}{cmtt}{bx}{n}{<5><6><7><8><9><10><10.95><12><14.4><17.28><20.74><24.88>cmttb10}{}

\lstdefinelanguage{smtlib2}{
  alsoletter=-,
  morekeywords={declare-const,define-fun,assert,minimize,maximize,check-sat,get-objectives,and,or,not,distinct},
  extendedchars=false,
  keywordstyle=\bfseries\color{lstpurple},
  deletekeywords={Int,Bool},
  keywords=[2]{Int,Bool},
  keywordstyle=[2]\color{lstblue},
}

\lstdefinestyle{psql}
{
tabsize=2,
basicstyle=\scriptsize\upshape\ttfamily,
language=SQL,
morekeywords={PROVENANCE,BASERELATION,INFLUENCE,COPY,ON,TRANSPROV,TRANSSQL,TRANSXML,CONTRIBUTION,COMPLETE,TRANSITIVE,NONTRANSITIVE,EXPLAIN,SQLTEXT,GRAPH,IS,ANNOT,THIS,XSLT,MAPPROV,cxpath,OF,TRANSACTION,SERIALIZABLE,COMMITTED,INSERT,INTO,WITH,SCN,UPDATED},
extendedchars=false,
keywordstyle=\bfseries,
mathescape=true,
escapechar=@,
sensitive=true
}

\lstdefinestyle{psqlcolor}
{
tabsize=2,
basicstyle=\scriptsize\upshape\ttfamily,
language=SQL,
morekeywords={PROVENANCE,BASERELATION,INFLUENCE,COPY,ON,TRANSPROV,TRANSSQL,TRANSXML,CONTRIBUTION,COMPLETE,TRANSITIVE,NONTRANSITIVE,EXPLAIN,SQLTEXT,GRAPH,IS,ANNOT,THIS,XSLT,MAPPROV,cxpath,OF,TRANSACTION,SERIALIZABLE,COMMITTED,INSERT,INTO,WITH,SCN,UPDATED},
extendedchars=false,
keywordstyle=\bfseries\color{lstpurple},
deletekeywords={count,min,max,avg,sum},
keywords=[2]{count,min,max,avg,sum},
keywordstyle=[2]\color{lstblue},
stringstyle=\color{lstreddark},
commentstyle=\color{lstgreen},
mathescape=true,
escapechar=@,
sensitive=true
}

\lstdefinestyle{datalog}
{
basicstyle=\footnotesize\upshape\ttfamily,
language=prolog
}

\lstdefinestyle{pseudocode}
{
  tabsize=3,
  basicstyle=\small,
  language=c,
  morekeywords={if,else,foreach,case,return,in,or},
  extendedchars=true,
  mathescape=true,
  literate={:=}{{$\gets$}}1 {<=}{{$\leq$}}1 {!=}{{$\neq$}}1 {append}{{$\listconcat$}}1 {calP}{{$\cal P$}}{2},
  keywordstyle=\color{lstpurple},
  escapechar=&,
  numbers=left,
  numberstyle=\color{lstgreen}\small\bfseries, 
  stepnumber=1, 
  numbersep=5pt,
}

\lstdefinestyle{xmlstyle}
{
  tabsize=3,
  basicstyle=\small,
  language=xml,
  extendedchars=true,
  mathescape=true,
  escapechar=£,
  tagstyle=\color{keywordpurple},
  usekeywordsintag=true,
  morekeywords={alias,name,id},
  keywordstyle=\color{lstred}
}

\lstdefinestyle{smtlib2}
{
tabsize=2,
basicstyle=\scriptsize\upshape\ttfamily,
numbers=left,
stepnumber=1,
breaklines=true,
stringstyle=\color{lstreddark},
commentstyle=\color{lstgreen},
mathescape=true,
escapechar=@,
sensitive=true
}

\begin{document}

\title{\ours: Efficient Batch Processing of Counting Queries}



\author{Yuxi Liu}
\affiliation{%
\institution{Duke University}
}
\email{yuxi.liu@duke.edu}

\author{Xiao Hu}
\affiliation{%
\institution{University of Waterloo}
}
\email{xiaohu@uwaterloo.ca}

\author{Pankaj K. Agarwal}
\affiliation{%
\institution{Duke University}
}
\email{pankaj@cs.duke.edu}

\author{Jun Yang}
\affiliation{%
\institution{Duke University}
}
\email{junyang@cs.duke.edu}

\begin{abstract}
\reva{Counting queries are ubiquitous in database systems, particularly for driving internal system optimization.}
Learned models for cardinality estimation rely heavily on large-scale training data,
yet generating such data by executing massive batches of counting queries is expensive.
We propose \ours, an efficient algorithm for batch evaluation of counting queries on top of a database system,
without modifying its internals.
\ours\ integrates the idea of factorized databases with a workload-aware domain quantization strategy,
allowing it to evaluate batches of counting queries using compact data structures rather than materializing massive join results.
\reva{\ours{}'s design is compatible with most database management system,
and we have implemented it as a client-side application on PostgreSQL with a lightweight C-language UDF (user-defined function).
This implementation delivers}
speedups between 2$\times$ and 178$\times$ over baselines
and good performance across various workloads,
making training and maintenance of learned cardinality estimation models significantly more practical.
\end{abstract}

\maketitle

\pagestyle{\vldbpagestyle}
\begingroup\small\noindent\raggedright\textbf{PVLDB Reference Format:}\\
\vldbauthors. \vldbtitle. PVLDB, \vldbvolume(\vldbissue): \vldbpages, \vldbyear.\\
\href{https://doi.org/\vldbdoi}{doi:\vldbdoi}
\endgroup
\begingroup
\renewcommand\thefootnote{}\footnote{\noindent
This work is licensed under the Creative Commons BY-NC-ND 4.0 International License. Visit \url{https://creativecommons.org/licenses/by-nc-nd/4.0/} to view a copy of this license. For any use beyond those covered by this license, obtain permission by emailing \href{mailto:info@vldb.org}{info@vldb.org}. Copyright is held by the owner/author(s). Publication rights licensed to the VLDB Endowment. \\
\raggedright Proceedings of the VLDB Endowment, Vol. \vldbvolume, No. \vldbissue\ %
ISSN 2150-8097. \\
\href{https://doi.org/\vldbdoi}{doi:\vldbdoi} \\
}\addtocounter{footnote}{-1}\endgroup

\ifdefempty{\vldbavailabilityurl}{}{
\vspace{.3cm}
\begingroup\small\noindent\raggedright\textbf{PVLDB Artifact Availability:}\\
The source code, data, and/or other artifacts have been made available at \url{\vldbavailabilityurl}.
\endgroup
}

\section{Introduction}
\label{sec:intro}

\common{Batches of} counting queries are not only useful in their own right for database applications,
but also frequently serve to collect basic statistics from data for monitoring and optimization.
With the growing popularity of \emph{learned query optimization}~\cite{10.1145/3626246.3654692, 10.1145/3448016.3452838, 10.14778/3342263.3342644, 10.1145/3588963} in recent years,
an interesting workload has emerged:
collecting training data for \emph{learned cardinality estimation} (CE)~\cite{park2020quicksel,10.14778/3329772.3329780,10.14778/3476249.3476254,wu2021unified,kipf2018learned,10.14778/3476249.3476259,DBLP:conf/edbt/MullerWL23,10.1145/3514221.3517896,jgmp}.
CE is a critical component of query optimization, as its accuracy directly impacts the quality of query execution plans~\cite{scheufele1997complexity, 10.14778/2850583.2850594}.
In these workloads, queries typically involve joins and selections over base tables but report only the final counts of the result sets.
These query-count pairs are subsequently used to train CE models.
Such queries may be derived from past workloads or synthesized specifically for training.
While the number of distinct join patterns is naturally limited by the database schema,
the queries themselves additionally contain a variety of selection conditions, targeting different attributes with varying constants.
Beyond simple equality comparisons, many conditions involve inequality or range predicates.
\Cref{tab:query_workloads} summarizes nine public query workloads widely used in CE research for training and evaluation.
These workloads span three commonly studied benchmark databases---IMDB~\cite{10.14778/2850583.2850594}, STATS~\cite{10.14778/3503585.3503586} and DSB~\cite{10.14778/3484224.3484234}---and capture a broad range of query shapes, join structures, and predicate types.

Executing such workloads is time-consuming.
Even when viewed as a one-time training cost, the overhead can be daunting;
in our experience, some workloads require hours or even more than a day to execute (\Cref{sec:experiment}).
Furthermore, models are rarely trained once and then forgotten.
As the database state changes, they can become outdated~\cite{10.1145/3514221.3526179,10.1145/3588713}:
cardinalities from previously executed queries may no longer be accurate, necessitating the re-execution of queries to retrain the models.
While monitoring real result counts from query execution feedback can mitigate this issue, simply observing final query result counts is insufficient~\cite{kipf2018learned,10.14778/3503585.3503586,jgmp}.
Effective training requires
(1)~the size of intermediate results for subqueries within \emph{potentially optimal} plans, not just the plan actually executed, and
(2)~feedback from unseen queries to prepare for potential workload shifts~\cite{10.14778/3725688.3725708,10.1145/3514221.3526179,10.1145/3639293}.
In a continuously running environment, executing these monitoring or training queries must not disrupt normal workloads.
These observations underscore the need for more efficient support for executing batches of counting queries.

Given the extensive literature on query processing, many ideas are applicable to the general problem of batch counting queries---%
including multi-query optimization, scalable continuous query processing, and factorized databases---%
see Section~\ref{sec:related} for more discussion.
A reasonable starting point, leveraging multi-query optimization, is to exploit the fact that many queries share the same join pattern,
differing only in their selection conditions.
Instead of executing them independently, one could perform the full join first and then apply individual selection conditions to obtain per-query counts.
While this baseline enables shared join processing, it leads to materializing the full join result, which can be massive for large databases.

We propose \emph{\textbf{\ours}}, a practical, scalable algorithm for efficient \emph{\textbf{Ba}tch processing of \textbf{Co}u\textbf{n}ting queries}.
A key idea, inspired by work on factorized databases~\cite{10.14778/2350229.2350242,DBLP:conf/icdt/OlteanuZ12,10.14778/2556549.2556579}, is that a counting query over a join can be evaluated without enumerating the join result.
While the count operator cannot generally be pushed down through a join, if the join is processed in a ``factorized'' manner---grouped by joining attribute values---we can simply count the joining ``factors'' and multiply these counts, avoiding enumeration of the joined result tuples.
A second key idea is workload-aware quantization of selection attribute domains.
This idea allows the data required for counting queries to be compressed, naturally exploiting the overlap and sharing of selection predicates within the workload.
\ours\ seamlessly combines these ideas, using compact ``count maps'' rather than bloated join tuples to represent intermediate results;
to combine these intermediate results, it uses a pair of operators $\otimes$ and $\oplus$, defined with clear semantics and amenable to optimization.
Finally, to ensure practicality and ease of adoption,
\reva{we implement \ours{} as a client application on top of a database system (DBMS) without modifying its internals.
\ours\ leverages the DBMS for processing,
utilizes user-defined functions (UDFs),
and carefully balances database- and client-side execution to minimize interface overhead.}

Our experimental results show that \ours{} is highly competitive for training workloads for learned CE in \Cref{tab:query_workloads}, achieving $2\times$ to $178\times$ speedups over baselines.
Moreover, \ours\ performs well across diverse workloads, providing significant speedups for expensive join patterns while remaining competitive on those where baseline approaches are already efficient.

\begin{table}[t]
    \caption{\mdseries Summary of nine query workloads across three databases, including the workload name with the link to file, associated database, number of queries (Qs), number of tables per query (Ts), number of join patterns (formally defined in \Cref{sec:prelim}), and a short description with source in the second row for each workload.}
    \label{tab:query_workloads}
    \vspace{-1em}
    \small
    \begin{tabular}{l|c|c|c|c}
        Query Workload & DB & \# Qs & \# Ts & \# join patterns
        \\\hline\hline
            \textbf{\texttt{synthetic}}~\cite{synthetic_sql} & IMDB & 5,000 & 1 to 3 & 16
        \\\cline{2-5}
            \multicolumn{5}{p{0.95\columnwidth}}{
                \footnotesize
                \reva{\citet{kipf2018learned}}: generated by the training data generator with a different seed
            } 
        \\\hline
            \textbf{\texttt{scale}}~\cite{scale_sql} & IMDB & 500 & 1 to 5 & 31 
        \\\cline{2-5}
            \multicolumn{5}{p{0.95\columnwidth}}{
                \footnotesize
                \reva{\citet{kipf2018learned}}: designed to show how MSCN generalizes to more joins.
            }
        \\\hline
            \textbf{\texttt{job-light}}~\cite{job_light_sql} & IMDB & 70 & 2 to 5 & 18 
        \\\cline{2-5}
            \multicolumn{5}{p{0.95\columnwidth}}{
                \footnotesize
                \reva{\citet{kipf2018learned}}: derived from JOB~\cite{10.14778/2850583.2850594}, excluding string predicates and disjunctions.
            }
        \\\hline
            \textbf{\texttt{job-light-single}}~\cite{job_light_single_sql} & IMDB & 254 & 1 & 1
        \\\cline{2-5}
            \multicolumn{5}{p{0.95\columnwidth}}{
                \footnotesize
                \reva{\citet{10.14778/3503585.3503586}}: single-table sub-plan queries extracted from job-light.
            }
        \\\hline
            \textbf{\texttt{job-light-join}}~\cite{job_light_join_sql} & IMDB & 696 & 2 to 5 & 27
        \\\cline{2-5}
            \multicolumn{5}{p{0.95\columnwidth}}{
                \footnotesize
                \reva{\citet{10.14778/3503585.3503586}}: join sub-plan queries extracted from job-light.
            }
        \\\hline
            \textbf{\texttt{stats-ceb}}~\cite{stats_ceb_sql} & STATS & 146 & 2 to 7 & 58
        \\\cline{2-5}
            \multicolumn{5}{p{0.95\columnwidth}}{
                \footnotesize
                \reva{\citet{10.14778/3503585.3503586}}: designed to be more comprehensive, with more diverse queries and more complex join patterns on STATS, compared to job-light on IMDB.
            }
        \\\hline
            \textbf{\texttt{stats-ceb-single}}~\cite{stats_ceb_single_sql} & STATS & 632 & 1 & 1
        \\\cline{2-5}
            \multicolumn{5}{p{0.95\columnwidth}}{
                \footnotesize
                \reva{\citet{10.14778/3503585.3503586}}: single-table sub-plan queries extracted from stats-ceb.
            }
        \\\hline
            \textbf{\texttt{stats-ceb-join}}~\cite{stats_ceb_join_sql} & STATS & 2,603 & 2 to 7 & 120
        \\\cline{2-5}
            \multicolumn{5}{p{0.95\columnwidth}}{
                \footnotesize
                \reva{\citet{10.14778/3503585.3503586}}: join sub-plan queries extracted from stats-ceb.
            }
        \\\hline
            \textbf{\texttt{dsb-grasp-20k}}~\cite{dsb_grasp_20k_sql} & DSB & 20,000 & 1 to 5 & 16
        \\\cline{2-5}
            \multicolumn{5}{p{0.95\columnwidth}}{
                \footnotesize
                A sample of the original workload~\cite{original_grasp_sql} generated by SeConCDF~\cite{10.14778/3725688.3725708} and GRASP~\cite{10.14778/3742728.3742745}
                on DSB. From 149,828 original queries, we remove queries not applicable to our methods and randomly sample 20,000 queries (details in \Cref{sec:experiment}).
            }
        \\\hline
    \end{tabular}
\end{table}

\section{Preliminaries}\label{sec:prelim}

\paragraph{Problem Statement and Notations}

Consider a database $\database$ with $n$ tables $\tab_1, \dots, \tab_n$,
where each table $\tab_i$ has a set of attributes denoted $\revc{\Attrs{\tab_i}}$.
We are interested in processing a set of counting queries $\revc{\queryset}$ over $\database$.%
\footnote{\common{For simplicity, we discuss only counting queries here, but our methods easily generalize to all \emph{distributive} and \emph{algebraic} SQL aggregates~\cite{jim_gray_data_cube} including also \sql{SUM}, \sql{MIN}, \sql{MAX},}
\common{\sql{AVG}, and \sql{STDDEV}, where each input expression is a single attribute or computable over attributes from the same table.}}
Each query counts the number of tuples returned by a single-table selection or a selection-join over a subset of the tables in $\database$.
We assume equality joins with no cycles or self-joins, and that the selection is a conjunction of predicates comparing a single attribute with a literal using \sql{=}, \sql{>}, \sql{<=}, etc.
We begin by introducing some useful notations \revc{(a table of notations is presented in the full version~\cite{fullversion})}.
Given each query $\query \in \revc{\queryset}$:
\begin{itemize}[leftmargin=*]
\item $\revc{\Tables{\query}}$ denotes the subset of tables in $\database$ referenced by $\query$.
\item $\revc{\JAttrs{\query}}$ denotes the subset of attributes in $\revc{\Tables{\query}}$ referenced by the join predicates of $\query$.
\item $\revc{\JPreds{\query}}$ denotes $\query$'s join predicates,
    represented as the set of equivalent classes of attributes in $\revc{\JAttrs{\query}}$ induced by $\query$'s join predicates.
    Two attributes $\attr_1, \attr_2$ belong to the same equivalent classes iff $\attr_1 = \attr_2$ is logically implied by $\query$'s join predicates.
\item Given a pair of disjoint subsets $\revc{\tabs}$ and $\revc{\tabs}'$ of tables in $\revc{\Tables{\query}}$,
    $\revc{\JAttrs{\query}}[\revc{\tabs}|\revc{\tabs}']$ denotes the subset of attributes of $\revc{\tabs}$ used by $\query$ to join with $\revc{\tabs}'$;
    i.e., $\revc{\JAttrs{\query}}[\revc{\tabs}|\revc{\tabs}'] = \{ \attr \in \bigcup_{\tab \in \revc{\tabs}} \revc{\Attrs{\tab}} \mid
        \exists \revc{\equivalentclass} \in \revc{\JPreds{\query}}:
        \attr \in \revc{\equivalentclass} \land (\exists \attr' \in \revc{\equivalentclass}: \attr' \in \bigcup_{\tab \in \revc{\tabs}'} \revc{\Attrs{\tab}}) \}.$
\item $\revc{\SAttrs{\query}}$ denotes the subset of attributes in $\revc{\Tables{\query}}$ referenced by $\query$'s selection predicates.
\item $\revc{\SPreds{\query}}$ denotes $\query$'s selection predicates,
    represented as a mapping from each attribute $\attr \in \revc{\SAttrs{\query}}$ to a range $G$ over the domain of $\attr$,
    such that $G$ is the widest range such that $\attr \in G$ is logically implied by $\query$'s selection predicates.
\item Given a subsets $\revc{\tabs}$ of tables in $\revc{\Tables{\query}}$,
    $\revc{\SAttrs{\query}}[\revc{\tabs}]$ denotes the set of attributes of $\revc{\tabs}$ referenced by $\query$'s selection conditions;
    i.e., $\revc{\SAttrs{\query}}[\revc{\tabs}] = \revc{\SAttrs{\query}} \cap \bigcup_{\tab \in \revc{\tabs}} \revc{\Attrs{\tab}}$.
\end{itemize}

For example, \Cref{sql:stats_ceb_example} shows the first four queries in the workload \texttt{stats\_ceb}~\cite{stats_ceb_sql} over the STATS database~\cite{10.14778/3503585.3503586}.
For $\query_4$:
\begin{itemize}[leftmargin=*]
\item $\Tables{\query_4} = \{ \sql{comments}, \sql{postHistory} \}$;
\item $\JAttrs{\query_4} = \{ \sql{comments.UserId}, \sql{postHistory.UserId} \}$;
\item $\JPreds{\query_4} = \{ \{ \sql{comments.UserId}, \sql{postHistory.UserId} \} \}$ (both join attributes are in one equivalence class);
\item $\JAttrs{\query_4}[\{\sql{comments}\}|\{\sql{postHistory}\}] = \{ \sql{comments.UserId} \}$;
\item $\JAttrs{\query_4}[\{\sql{postHistory}\}|\{\sql{comments}\}] = \{ \sql{postHistory.UserId} \}$;
\item $\SAttrs{\query_4} = \{ \sql{postHistory.PostHistoryTypeId}, \sql{postHistory.CreationDate} \}$;
\item $\SPreds{\query_4} = \{ \sql{postHistory.PostHistoryTypeId} \mapsto [1, 1],\\
\sql{postHistory.CreationDate} \mapsto [\sql{'2010-09-14 11:59:07'::timestamp}, \infty) \}$.
\end{itemize}

\begin{sqllisting}[float=t]{\mdseries First 4 queries in \texttt{stats\_ceb}~\cite{stats_ceb_sql}.}{sql:stats_ceb_example}
SELECT COUNT(*) FROM badges as b, users as u -- $\query_1$
 WHERE b.UserId = u.Id AND u.UpVotes >= 0;
SELECT COUNT(*) FROM comments as c, badges as b -- $\query_2$
 WHERE c.UserId = b.UserId AND c.Score = 0 AND b.Date <= '2014-09-11 14:33:06'::timestamp;
SELECT COUNT(*) FROM comments as c, postHistory as ph -- $\query_3$
 WHERE c.UserId = ph.UserId AND c.Score = 0 AND ph.PostHistoryTypeId = 1;
SELECT COUNT(*) FROM comments as c, postHistory as ph -- $\query_4$
 WHERE c.UserId = ph.UserId AND ph.PostHistoryTypeId = 1 AND ph.CreationDate >= '2010-09-14 11:59:07'::timestamp;
\end{sqllisting}

\paragraph{The First Baseline: Independent Processing (\sequentialproc)}

A straightforward solution, adopted by most existing work on learned CE, runs the queries in $\revc{\queryset}$ one by one, independently,
using the DBMS that manages $\database$.
We call this approach \sequentialproc{} for short.
The performance of \sequentialproc\ is heavily dependent on the underlying DBMS.
A capable DBMS will optimize each query by pushing down selection conditions, reordering joins, and choosing appropriate join methods,
leveraging data statistics and available indices.
Caching by the DBMS buffer pool may also improve execution performance across queries.
On the other hand, most DBMS lack advanced methods for optimizing multiple queries simultaneously.
Moreover, most of them choose to execute a counting query $\query$ by first joining all tables in $\query$ before applying the final \sql{COUNT} aggregation, failing to explore opportunities for pushing \sql{COUNT} below joins.

\paragraph{Join Patterns}

Before presenting a more advanced baseline solution as well as our solution in \Cref{sec:ours},
we introduce a concept used by both.
The \emph{join pattern} of a query $\query$ is characterized by $\langle \revc{\Tables{\query}}, \revc{\JPreds{\query}} \rangle$, i.e.,
the tables that $\query$ joins, and $\query$'s join predicates.
Suppose queries in $\revc{\queryset}$ have $\revc{\njps}$ distinct join patterns $\revc{\jp}_1, \ldots, \revc{\jp}_\revc{\njps}$;
these patterns partition $\revc{\queryset}$ into a disjoint union of $\revc{\njps}$ subsets of queries,
denoted $\revc{\queryset}[\revc{\jp}_1], \ldots, \revc{\queryset}[\revc{\jp}_\revc{\njps}]$,
where each $\revc{\queryset}[\revc{\jp}_i]$ is the set of queries with join pattern $\revc{\jp}_i$
(but may differ in their selection predicates). For example, $\query_3$ and $\query_4$ in \Cref{sql:stats_ceb_example} have the same join pattern
(with tables $\{ \sql{comments}, \sql{postHistory} \}$ and join predicate \sql{comments.UserId} \sql{=} \sql{postHistory.UserId},
despite having different selection predicates).
$\query_1$ and $\query_2$ each contribute a distinct join pattern.
In practice, because all queries in $\revc{\queryset}$ come from the same underlying database $\database$,
the number of distinct join patterns tends to be much smaller than the number of queries.
For example, JOB's \texttt{synthetic}~\cite{synthetic_sql} contains 5{,}000 queries but only 16 distinct join patterns.

\paragraph{The Second Baseline: Join and Post-Filtering (\postfilter)}

This approach seeks to avoid redundant computation across queries with identical join patterns.
We pre-process the queries into partitions $\revc{\queryset}[\revc{\jp}_1], \ldots, \revc{\queryset}[\revc{\jp}_\revc{\njps}]$ according to their join patterns.
This step requires only a scan over $\revc{\queryset}$ to perform syntactical analysis.
For each partition $\revc{\queryset}[\revc{\jp}_i]$, we compute the join only once,
followed by a post-filtering step,
which checks each join result tuple against the selection predicates of queries in $\revc{\queryset}[\revc{\jp}_i]$
to determine which queries' counts to increment.
We call this baseline \postfilter.

There are many options for implementing the post-filtering step, as discussed in \Cref{sec:related},
including methods that first build a data structure for $\revc{\queryset}[\revc{\jp}_i]$
for efficient identification and updating of counters in time sublinear \revc{to} $|\revc{\queryset}[\revc{\jp}_i]|$ per join result tuple.
While these methods can scale to thousands or millions of queries,
the typical workloads we target do not have so many queries per join pattern, as evidenced in \Cref{tab:query_workloads}.
Through our experiments, we have found a simple strategy leveraging the underlying DBMS to be the most effective.
Given a join pattern $\revc{\jp} = \langle \Tables{\revc{\jp}}, \JPreds{\revc{\jp}} \rangle$ and $k$ queries $\query_1, \ldots, \query_k$ sharing this pattern, we issue a single SQL query in \Cref{sql:postfilter} to compute all of them.
Here, $\SPreds{}$, $\Tables{}$, and $\JPreds{}$ are translated into SQL.
Each \sql{CASE} expression determines whether a join result tuple contributes to a query $\query_i$
by checking its selection predicates (a failed check yields \sql{NULL}, which is ignored by \sql{COUNT}).
In the end, a single $k$-component result tuple is computed, with each component holding the result count for one query.

\begin{sqllisting}[float=t]{\mdseries Computing $k$ queries with join pattern $\revc{\jp}$ in \postfilter.}{sql:postfilter}
SELECT COUNT(CASE WHEN $\revc{\SPreds{\query_1}}$ THEN 1 END), ...,
    COUNT(CASE WHEN $\revc{\SPreds{\query_k}}$ THEN 1 END)
FROM $\revc{\Tables{\jp}}$ WHERE $\revc{\JPreds{\jp}}$;
\end{sqllisting}

\paragraph{Discussion}

Compared with \sequentialproc, \postfilter\ eliminates redundant computation of joins among queries sharing the same join pattern.
However, \sequentialproc\ can still outperform \postfilter\ if the selection predicates of these queries have little overlap,
and if database indices enable \sequentialproc\ to apply selective selection predicates early in query processing.
In contrast, \postfilter\ has no effective means to push selection predicates down
because in most workloads, the disjunction of all selection predicates from multiple queries
cannot be expressed as a succinct, ``sargable''~\cite{DBLP:conf/sigmod/SelingerACLP79} \sql{WHERE} condition without introducing many false positives.
Furthermore, most database optimizers do not push aggregation below joins, let alone those with conditional expression inputs as in \Cref{sql:postfilter}.
Therefore, \postfilter\ effectively enumerates the full join result before post-filtering and counting---a key limitation that we seek to overcome.

\section{Basic \ours}
\label{sec:ours}

This section introduces basic \ours, focusing on key ideas and the high-level algorithm.
Many implementation and optimization details are also crucial to making \ours\ competitive in practice,
but to simplify presentation, we defer them to \Cref{sec:implementation}.
We start with three key ideas, along with some results and definitions based on them;
we then describe the algorithm.
Like \postfilter, given a set $\revc{\revc{\queryset}}$ of queries, \ours\ partitions $\revc{\queryset}$ into subsets according to join patterns,
such that queries in each subset $\revc{\queryset}[\revc{\jp}]$ share the same join pattern $\revc{\jp} = \langle \revc{\Tables{\jp}}, \revc{\JPreds{\jp}} \rangle$.
Hence, most of this section focuses on how to process one such subset of queries given $\revc{\jp}$.

Before proceeding, we briefly present a geometric view of the problem for intuition.
Each result tuple in the cross product of $\revc{\Tables{\jp}}$ can be seen as a point in a high-dimensional space $\mathbb{S}$,
where each dimension corresponds to an attribute in $\bigcup_{\tab \in \revc{\Tables{\jp}}} \revc{\Attrs{\tab}}$,
ignoring those not referenced by any predicate in $\revc{\queryset}[\revc{\jp}]$.
Let $J$ denote the result of the full join of $\revc{\Tables{\jp}}$ using $\revc{\JPreds{\jp}}$.
Points in $J$ are those cross-product points that lie on a hyperplane $\mathbb{J}$ in $\mathbb{S}$ defined by $\revc{\JPreds{\jp}}$.
Each query $\query \in \revc{\queryset}[\revc{\jp}]$ corresponds to an orthogonal hyperrectangle in $\mathbb{S}$ with range predicates restricted to dimensions in $\SAttrs{\query}$.
Our problem boils down to counting, for each $\query$, how many points in $J$ fall into $\query$'s hyperrectangle.

Intuitively, these points are not positioned arbitrarily:
not only do they lie on $\mathbb{J}$ because of the join predicates,
but they also come from the cross product of $\revc{\Tables{\jp}}$, meaning that their projections onto the subspace $\mathbb{S}[\revc{\Attrs{\tab}}]$ for each $\tab \in \revc{\Tables{\jp}}$ cannot exceed $|\tab|$ distinct points, which is often much smaller than $|J|$.
\common{In fact, if we consider the subset of $J$ restricted to any particular combination of values for all join attributes,
their projections onto $\mathbb{S}[\bigcup_{\tab \in \revc{\Tables{\jp}}} \revc{\Attrs{\tab}}]$ will form a perfect Cartesian product
of their projections onto each of $\mathbb{S}[\revc{\Attrs{\tab}}]$ for $\tab \in \revc{\Tables{\jp}}$
(which we shall substantiate in \Cref{sec:ours:orthogonality}).
This property, along with the fact the query hyperrectangles may overlap significantly,}
makes it possible to perform counting tasks more efficiently than simply enumerating $J$ upfront (which $\postfilter$ does).

\common{To further exploit the overlaps among the query hyperrectangles,
consider a subset of $J$ that forms a Cartesian product described above.
We can partition its projection onto to each subspace $\mathbb{S}[\revc{\Attrs{\tab}}]$ into a coarse grid,
using the endpoints of the query range predicates
(a process we call ``quantization'' later in \Cref{sec:ours:quantization}).
For example, in \Cref{fig:quantization}, this projection consists of a collection of red points, partitioned by a $4 \times 4$ grid.
A key property of this grid is that all points within a cell lie in the same subset of query hyperrectangles.
Therefore, we can compress the red points in each cell to a single ``weighted'' point, yielding a compressed representation of $J$
(which we refer to as a ``count map'' later in \Cref{sec:ours:count-maps}).
This compressed representation is constructed recursively for each $\jp$, without having to enumerate $J$ first (\Cref{sec:ours:algo}).}

\subsection{Conditional Orthogonality of Joins}
\label{sec:ours:orthogonality}

The first idea has its roots in the well-studied problem of factorized databases and related \revb{worst-case optimal} join algorithms (further discussed in \Cref{sec:related}).
A simple observation is that, in order to count the number of tuples in a cross product between two tables $R_1$ and $R_2$,
we just need to calculate $|R_1| \times |R_2|$, without enumerating $R_1 \times R_2$.
While the same observation no longer holds for $|R_1 \join_{R_1.A=R_2.A} R_2|$ when join predicate exists, if we additionally set the join attribute value $A=\revc{x}$,
then the number of result tuples \emph{conditioned on this specific setting} can still be computed directly:
i.e., $|\selection_{A=\revc{x}} (R_1 \join_{R_1.A=R_2.A} R_2)| = |\selection_{A=\revc{x}} R_1| \times |\selection_{A=\revc{x}} R_2|$.
We can generalize this idea further to a star-shaped join as follows.

\begin{lemma}[Conditional Orthogonality of Joins]\label{lemma:orthogonality}
Consider a star-shaped join query centered at $E_0$:
\[ E_0 \join_{\theta_1} E_1 \join_{\theta_2} \cdots \join_{\theta_n} E_n. \]
Here, the $E_i$'s are subqueries, and for each $i = 1, \ldots, n$,
$\theta_i$ is a conjunctive predicate equating pairs of attributes from $E_0$ and $E_i$.
Note that there are no join predicates across $E_1, \ldots, E_n$.
Denote by $\revc{\attrs}_0$ the set of join attributes from $E_0$ referenced by $\theta_1, \ldots, \theta_n$.
Let $v$ be any mapping of every attribute $\attr \in \revc{\attrs}_0$ to a value $v(\attr)$ in $\attr$'s domain,
and let $v\llbracket\theta_i\rrbracket$ denote the condition obtained by applying $v$ to $\theta_i$
(i.e., replacing each $\attr \in \revc{\attrs}_0$ by $v(\attr)$---note that the resulting condition becomes a selection over $E_i$).
The following equivalence holds:
\begin{align*}
&\phantom{{}={}} \selection_{\land_{\attr \in \revc{\attrs}_0} \attr = v(\attr)} \left( E_0 \join_{\theta_1} E_1 \join_{\theta_2} \cdots \join_{\theta_n} E_n \right) \\
&= \left(\selection_{\land_{\attr \in \revc{\attrs}_0} \attr = v(\attr)} E_0\right)
\times \left(\selection_{v\llbracket\theta_1\rrbracket} E_1\right)
\times \cdots
\times \left(\selection_{v\llbracket\theta_n\rrbracket} E_n\right).
\end{align*}
\end{lemma}

\begin{figure}[t]
    \centering
    \includegraphics[width=0.7\columnwidth]{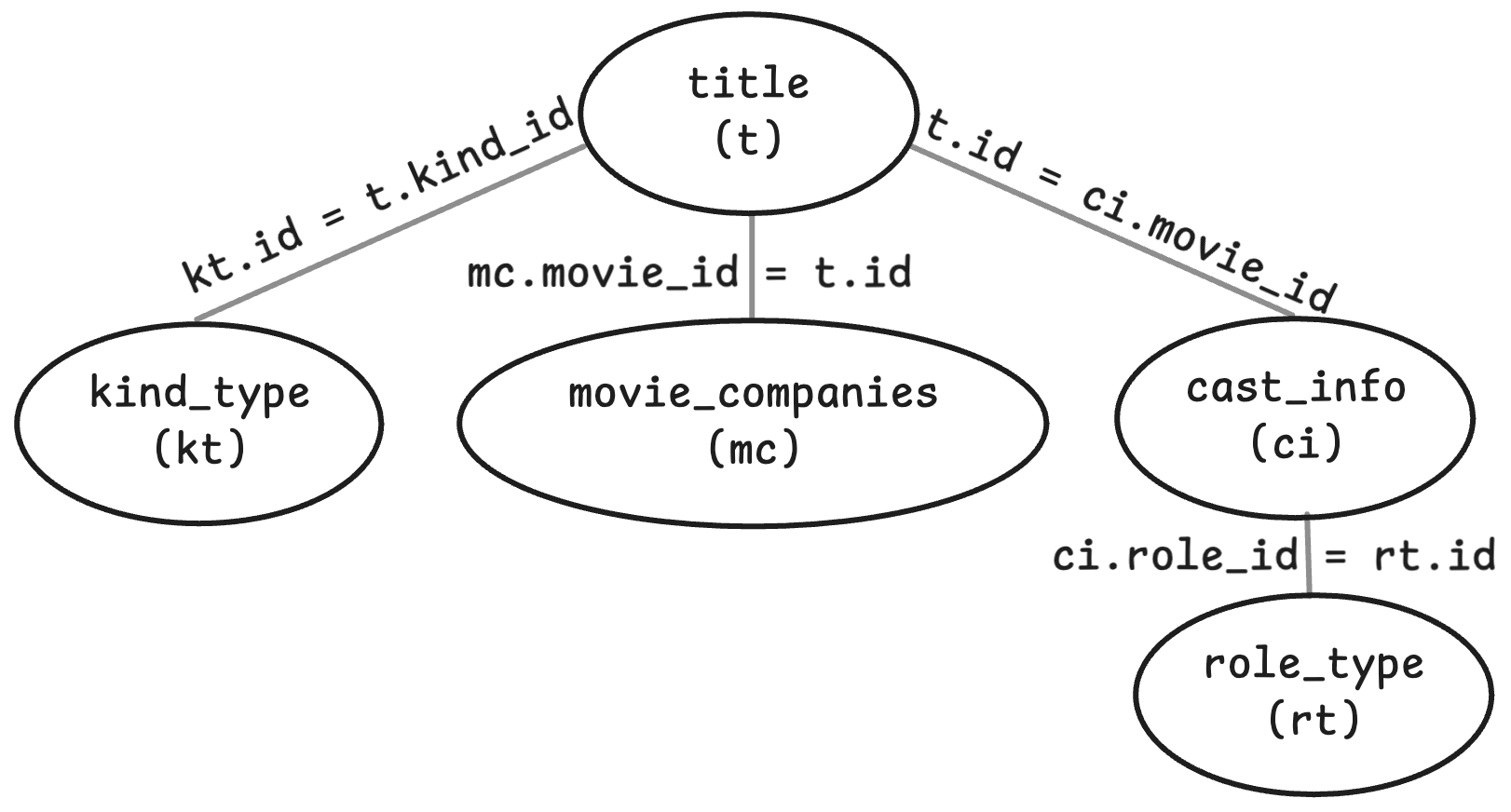}
    \vspace*{-2ex}
    \caption{\mdseries A join pattern in the IMDB schema~\cite{imdb_job_schema}.}
    \label{fig:join_pattern_tree}
\end{figure}

\begin{example}\label{eg:orthogonality}
Consider the join pattern in \Cref{fig:join_pattern_tree}.
Tables \sql{title AS t}, \sql{kind\_type AS kt}, and \sql{movie\_companies AS mc} correspond to $E_0, E_1, E_2$,  
while the join between \sql{cast\_info AS ci} and \sql{role\_type AS rt} corresponds to $E_3$.
In \sql{title}, suppose we fix \sql{t.kind\_id=}$x$ (which joins \sql{kt.id}) and \sql{t.id=}$y$ (which joins \sql{mc.movie\_id} and $\sql{ci}\join\sql{rt}$).
By \Cref{lemma:orthogonality},
the number of full join result tuples with $(\sql{t.kind\_id}, \sql{t.id}) = (x, y)$ can be computed directly as:
$|\selection_{\sql{t.kind\_id}=x \land \sql{t.id}=y} \sql{t}|
\times |\selection_{\sql{kt.id}=x} \sql{kt}|
\times |\selection_{\sql{mc.movie\_id}=y} \sql{mc}|
\times |\selection_{\sql{ci.movie\_id}=y} (\sql{ci} \join \sql{rt})|$,
without enumerating the full join result tuples.
To compute the full join result size overall,
we iterate over all possible combinations of $(\sql{t.kind\_id}, \sql{t.id})$ values in \sql{t},
apply the above procedure to each combination,
and tally the total.
\end{example}

The style of processing illustrated by the above example has been used recently for efficient computation of aggregate queries~\cite{10.14778/3718057.3718068}.
\common{The subexpression $|\selection_{\sql{ci.movie\_id}=y} (\sql{ci} \join \sql{rt})|$} can be processed by the same procedure.
Later in this section, we will see how to extend the idea to computing result counts of multiple queries with different selection predicates beyond simply counting the full join result.

\subsection{Quantization of Selection Attributes}
\label{sec:ours:quantization}

While tables and attribute domains can be large,
the number of queries in $\revc{\queryset}$ places a natural constraint on the number of constants from each domain appearing in selection predicates.
In other words, from the perspective of $\revc{\queryset}$, fine-grained differences among attribute values may not affect result counts.
Our key idea is to compress the attribute domains using workload-aware \emph{quantiziation},
turning large, complex domains into a small range of integers that are efficient to work with.

\revc{Given a join pattern $\revc{\jp}$,}
let $\revc{\SAttrs{\jp}} = \bigcup_{\query \in \revc{\queryset}[\revc{\jp}]} \SAttrs{\query}$ denote the set of selection attributes in all queries of pattern $\revc{\jp}$.
We construct a \emph{quantization scale} $\QuantScale_\attr$ for each selection attribute $\attr \in \revc{\SAttrs{\jp}}$ as follows. 
First, we extract from $\revc{\queryset}[\revc{\jp}]$ the set $\revc{\SPreds{\jp}(\attr)} = \{ \SPreds{\query}(\attr) \mid \exists \query \in \revc{\queryset}[\revc{\jp}]: \attr \in \SPreds{\query} \}$ 
of predicate ranges associated with $\attr$.
We sort all range endpoints, which partition the domain of $\attr$ into an ordered list of atomic ranges.
For each atomic range contained in at least one query range in $\revc{\SPreds{\jp}(\attr)}$, we create a new \emph{bucket} in $\QuantScale_\attr$ and assign it a serial integer id (starting with $1$).
Hence, $\QuantScale_\attr$ maps each relevant atomic range (bucket) to an integer, preserving order.
A value $x$ from $\attr$'s domain is quantized into an integer $\QuantScale_\attr(x)$, the id of the bucket containing $x$,
or $0$ if $x$ lies outside all of buckets in $\QuantScale_\attr$.
\reva{We present the detailed construction algorithm in the full version~\cite{fullversion}},
which handles additional intricacies with open or close intervals.

Let $\QuantScales_\revc{\jp} = \{ \QuantScale_A \mid A \in \revc{\SAttrs{\jp}} \}$ denote the collection of all attribute quantization scales for queries in  $\revc{\queryset}[\revc{\jp}]$,
and $\QuantScales_\revc{\jp}[\tab] = \{ \QuantScale_A \mid A \in \revc{\SAttrs{\jp}} \cap \revc{\Attrs{\tab}} \}$ denote those for attributes in table $\tab \in \revc{\Tables{\jp}}$.
\common{Returning to the geometric view introduced at the beginning of the section,
$\QuantScales_\revc{\jp}[\tab]$ induces a grid over the subspace of $\mathbb{S}$ spanning the selection attribute dimensions of $\tab$.
This grid allows us to map each tuple in $R$, by its selection attribute values $\langle x_1, \ldots, x_k \rangle$,
to a \emph{grid coordinate} $\myvec{b} = (b_1, \ldots, b_k)$,
where each $b_i = \QuantScale_{\attr_i}(x_i)$ is the bucket id for value $x_i$ in the quantization scale for attribute $\attr_i$.}

\common{Continuing with the geometric intuition,
collectively, quantization scales $\QuantScales_\revc{\jp}$ induces a grid over the subspace $\mathbb{S}[\revc{\SAttrs{\jp}}]$ consisting of the selection attribute dimensions across all tables in $\Tables{\jp}$.
By construction of $\QuantScales_\revc{\jp}$, all query hyperrectangles in $\queryset[\jp]$ perfectly align with grid boundaries.
Hence, all selection predicates can be quantized using the same scales.
Lemma \ref{lemma:quantization} below} formalizes the guarantee that precise evaluation of selection predicates is possible in the quantized space.
\begin{lemma}[Quantization Preserves Selections]\label{lemma:quantization}
    Given a set of selection-join queries $\queryset[\jp]$ and quantization scales $\QuantScales_\jp$ constructed from $\queryset[\jp]$,
    there exists a function $f(\QuantScale_\attr, \delta)$
    returning an integer range $[i_1, i_2]$ for a range $\delta$ over an attribute $\attr$ with quantization scale $\QuantScale_\attr \in \QuantScales_\jp$,
    such that for any $\query \in \queryset[\jp]$ and every selection predicate $\attr \mapsto \delta$ in $\SPreds{\query}$:
    $x \in \delta \Leftrightarrow \QuantScale_\attr(x) \in f(\QuantScale_\attr, \delta)$.
\end{lemma}

\begin{figure}[!t]
    \centering\includegraphics[width=1\columnwidth]{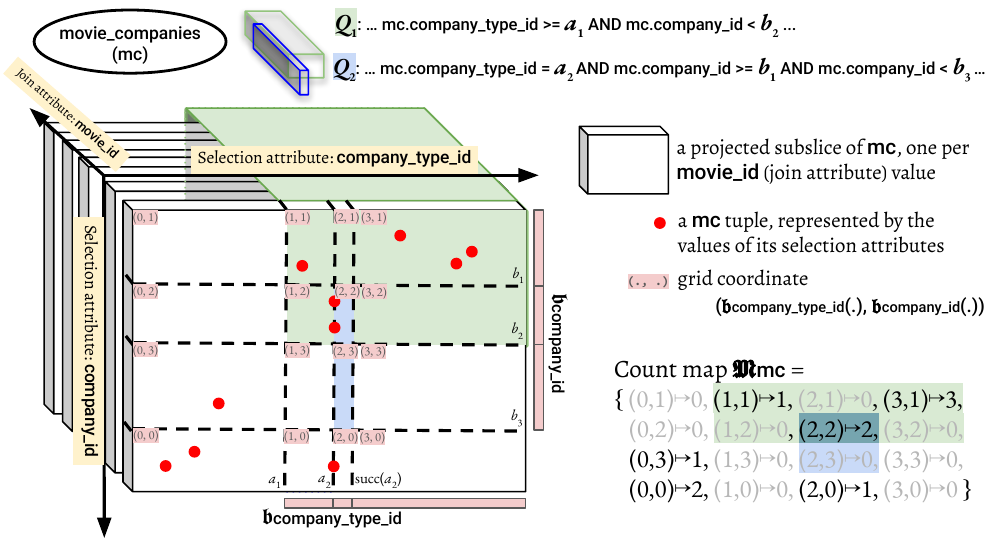}
    \vspace*{-4ex}
    \caption{\mdseries \common{Quantized selection count map for a projected subslice of table \sql{mc}.
    The quantization scales for \sql{mc} are induced by queries $\query_{1}$ and $\query_{2}$ (only selection predicates on \sql{mc} are shown).
    }}
    \label{fig:quantization}
\end{figure}

\common{As an example, \Cref{fig:quantization} shows the grid over table $\sql{mc}$ induced by two queries.
The quantization scales $\QuantScale_{\sql{company\_type\_id}}$ and $\QuantScale_{\sql{company\_id}}$ respectively correspond to
the two $\sql{mc}$ attributes referenced by the queries' selection predicates.
(For now, ignore the mention of ``projected subslice,'' which will be formally introduced in \Cref{sec:ours:algo}.)
The buckets for $\QuantScale_{\sql{company\_type\_id}}$, numbered $1$ through $3$, are $[a_1,a_2)$, $[a_2,\narrow{succ}(a_2))$, and $[\narrow{succ}(a_2), \infty)$,
where $\narrow{succ}(a_2)$ denotes the successor value of $a_2$ in the domain;
any value in $(-\infty, a_1)$ will be quantized to special bucket id $0$ because this range is not contained in any selection predicate.
The buckets for $\QuantScale_{\sql{company\_id}}$ are similarly induced by selection predicates involving $\sql{company\_id}$.}

\subsection{Quantized Selection Count Maps}
\label{sec:ours:count-maps}
Since queries in $\revc{\queryset}[\revc{\jp}]$ ultimately only care about counts, a natural idea following selection attribute quantization is to further aggregate the points that fall into each grid cell induced by $\QuantScales_\revc{\jp}$ into a single count, instead of enumerating them.

Given a table $\tab \in \revc{\Tables{\jp}}$ with selection attributes $\{ \attr_1, \ldots, \attr_k \}$ and quantization scales $\QuantScales_\revc{\jp}[\tab] = \{ \QuantScale_{\attr_1}, \ldots, \QuantScale_{\attr_k} \}$,
we compress a subset of tuples in $\tab$ into a \emph{(quantized selection) count map} $\CountMap$:
each entry of $\CountMap$ maps a grid coordinate $\myvec{b} = (b_1, \ldots, b_k)$
to $\CountMap[\myvec{b}]$, the count of tuples within the grid cell---i.e.,
any tuple $t$ satisfying $\QuantScale_{\attr_i}(t.\attr_i) = b_i$ for $i = 1, \ldots, k$.
For example, \Cref{fig:quantization} shows the count map for the set of \common{tuples (2D points in the geometric view)}.
%

We generalize the concept of count map $\CountMap$ over any subset of tables $\revc{\tabs} \subseteq \revc{\Tables{\jp}}$.
Let $\revc{\SAttrs{\jp}[\tabs]} = \SAttrs{\revc{\jp}} \cap \bigcup_{\tab \in \revc{\tabs}} \revc{\Attrs{\tab}}$ denote the set of selection attributes in $\revc{\tabs}$ \revc{from all queries of pattern $\revc{\jp}$}.
A grid coordinate for $\CountMap$, with one component for each attribute in $\revc{\SAttrs{\tabs}}$,
identifies a grid cell induced by quantization scales $\{ \QuantScale_A \mid A \in \revc{\SAttrs{\tabs}} \}$
in subspace $\mathbb{S}[\revc{\SAttrs{\tabs}}]$.
Given a subset $T$ of tuples in the cross product of $\revc{\tabs}$ corresponding to points in $\mathbb{S}[\revc{\SAttrs{\tabs}}]$,
$\CountMap$ counts the corresponding points of $T$ in each grid cell.

Our goal is to construct a count map over the entire $\revc{\Tables{\jp}}$ for the full join result set $J$, but importantly, without enumerating $J$.
Intuitively, count maps represent intermediate results compactly. 
Obtaining a count map for a single table is easy, but combining count maps for different subsets of tables into a big one is complicated by joins,
because not all points in two grid cells in orthogonal subspaces join with each other.
However, leveraging \Cref{lemma:orthogonality}, we can process points in $J$ in groups:
points in each group all share appropriate join attribute values,
thereby allowing count maps to be computed for different subsets of tables and then conveniently ``multiplied.''
Then, the product count maps across groups can be ``added'' to obtain the final result count map.
An example will be provided in \Cref{sec:ours:algo}.

We formally define ``multiply'' and ``add'' as follows:
\begin{description}[leftmargin=1em]
\item[$\otimes$ (multiply)]
    $\CountMap_1$ and $\CountMap_2$ are count maps over disjoint subsets $\revc{\tabs}_1$ and $\revc{\tabs}_2$ of tables in $\revc{\Tables{\jp}}$.
    We define $\CountMap_1 \otimes \CountMap_2$, a count map over $\revc{\tabs}_1 \cup \revc{\tabs}_2$, as
    $\smash{\big\{ \myvec{b}_1\concat\myvec{b}_2 \mapsto (\CountMap_1[\myvec{b}_1]\cdot\CountMap_2[\myvec{b}_2]) \mid \myvec{b}_1 \in \CountMap_1, \myvec{b}_2 \in \CountMap \big\}}$, where $\concat$ concatenates grid coordinate vectors.
\item[$\oplus$ (add)]
    $\CountMap_1$ and $\CountMap_2$ are count maps over the same subset of tables $\revc{\tabs} \subseteq \revc{\Tables{\jp}}$
    (and thus same quantization scales).
    We define $\CountMap_1 \oplus \CountMap_2$, a count map over $\revc{\tabs}$, as
    $\smash{\big\{ \myvec{b} \mapsto (\CountMap_1[\myvec{b}]+\CountMap_2[\myvec{b}]) \mid \myvec{b} \in }
    $ $\smash{\CountMap_1 \big\}}$.
\end{description}

The following lemmas establish the correctness of using $\otimes$ and $\oplus$ for computing count maps.
Formally, given $\revc{\queryset}[\revc{\jp}]$, quantization scales $\QuantScales_\revc{\jp}$, a subset of tables $\revc{\tabs} \subseteq \revc{\Tables{\jp}}$,
and a subset $T$ of tuples in the cross product of $\revc{\tabs}$,
we say that a count map $\CountMap$ over $\revc{\tabs}$ is \emph{complete} with respect to $T$
if for any query $\query \in \revc{\queryset}[\revc{\jp}]$,
the size of the intersection between $T$ and the selection-join subquery of $\query$ restricted%
\footnote{More precisely, we remove from $\query$ any table outside $\revc{\tabs}$;
    remove from any equivalent class in $\JPreds{\query}$ any attribute outside $\revc{\tabs}$ and then remove any empty or singleton equivalent class;
    and remove any entry in $\SPreds{\query}$ for an attribute outside $\revc{\tabs}$.}
to $\revc{\tabs}$ can be computed from $\CountMap$ and $\QuantScales_\revc{\jp}$.

\begin{lemma}[Multiplying Count Maps]\label{lemma:mutiply-count}
Given $\revc{\queryset}[\revc{\jp}]$, consider disjoint subsets $\revc{\tabs}_0, \ldots, \revc{\tabs}_n$ of tables in $\revc{\Tables{\jp}}$,
where $\revc{\JPreds{\jp}}$ implies a join condition $\theta_i$ relating $\revc{\tabs}_0$ to $\revc{\tabs}_i$ for each $i = 1,\ldots,n$,
but there is no join condition across $\revc{\tabs}_1, \ldots, \revc{\tabs}_n$ that is not already implied by $\bigwedge_i \theta_i$.
For each $i = 1, \ldots, n$, let $E_i$ denote the join subquery of tables in $\revc{\tabs}_i$, with condition implied by $\revc{\JPreds{\jp}}$.
Consider the star-shaped join query centered at $E_0$:
\[ E_0 \join_{\theta_1} E_1 \join_{\theta_2} \cdots \join_{\theta_n} E_n, \]
which conforms to the query structure in \Cref{lemma:orthogonality}.
Denote by $\revc{\revc{\attrs}_0 = \JAttrs{\jp}[\tabs_0\,|\,\cup_{i\in[1,n]} \tabs_i]}$ the set of join attributes from $E_0$ referenced by $\theta_1, \ldots, \theta_n$.
Let $v$ denote any mapping of every attribute $\attr \in \revc{\attrs}$ to a value $v(\attr)$ in $\attr$'s domain,
for some attribute set $\revc{\attrs}$ where $\revc{\attrs}_0 \subseteq \revc{\attrs} \subseteq \bigcup_{\tab \in \revc{\tabs}_0} \revc{\Attrs{\tab}}$.
Suppose:
\begin{itemize}[leftmargin=*]
\item $\CountMap_0$ is a count map over $E_0$ complete w.r.t.\ $\selection_{\land_{\attr \in \revc{\attrs}} \attr = v(\attr)} E_0$; and
\item $\forall i = 1,\ldots,n$: $\CountMap_i$ is a count map over $E_i$ complete w.r.t.\ $\selection_{v\llbracket\theta_1\rrbracket} E_i$.
\end{itemize}
Then, $\CountMap_0 \otimes \CountMap_1 \otimes \cdots \otimes \CountMap_n$ is complete with respect to
\[ \selection_{\land_{\attr \in \revc{\attrs}} \attr = v(\attr)} \left( E_0 \join_{\theta_1} E_1 \join_{\theta_2} \cdots \join_{\theta_n} E_n \right). \]
\end{lemma}

\begin{lemma}[Adding Count Maps]\label{lemma:add-count}
Given $\revc{\queryset}[\revc{\jp}]$ and a subset of tables $\revc{\tabs} \subseteq \revc{\Tables{\jp}}$,
suppose $\CountMap_1, \ldots, \CountMap_n$ are count maps over $\revc{\tabs}$,
where each $\CountMap_i$ is complete with respect to a subset $T_i$ of tuples in the cross product of $\revc{\tabs}$.
If $T_1, \ldots, T_n$ are disjoint, then
$\CountMap_1 \oplus \cdots \oplus \CountMap_n$ is complete with respect to $\bigcup_{i \in [1,n]} T_i$.
\end{lemma}

\subsection{Basic \ours\ Algorithm}
\label{sec:ours:algo}

\begin{algorithm}[t]
\caption{$\ours(\revc{\queryset}, \database)$}\label{alg:our_main}
\begin{algorithmic}[1]
\Require{A set $\revc{\queryset}$ of counting selection-join queries over $\database$.}
\Ensure{Result count for each query in $\revc{\queryset}$.}
\State Scan $\revc{\queryset}{}$ and partition it into $\revc{\bigcup_\revc{\jp} \queryset[\jp]}$ by join pattern;
\For{each subset $\revc{\queryset[\jp]}$ of $\revc{\queryset}$}
    \State $\QuantScales_{\revc{\jp}} \gets \{ \ConstructQuantScales(\tab, \revc{\queryset[\jp]}) \mid \tab \in \Tables{\revc{\jp}} \}$;
    \State Determine the plan tree for $\revc{\jp}$;
    \State $\CountMap \gets \ourrecursive(\revc{\treeroot(\jp)}, \emptyset)$;\label{alg:our_main:call_recurisve}
    \State \textbf{yield} \textbf{from} $\ComputeCounts(\revc{\queryset[\jp]}, \CountMap)$;
\EndFor
\end{algorithmic}
\end{algorithm}

We are now ready to describe the overall \ours\ algorithm (\Cref{alg:our_main}).
First, we make a pass over all queries and partition them into subset where each contains queries with the same join pattern.
For each subset $\revc{\queryset}[\revc{\jp}]$ with join pattern $\revc{\jp}$, we construct the quantization scales for them, as discussed in \Cref{sec:ours:quantization}.
We then determine an ``execution plan'' for $\revc{\jp}$ as an ordered tree,
akin to the notion of ``(paths in) f-tree'' in factorized database literature~\cite{DBLP:conf/icdt/OlteanuZ12,10.14778/2350229.2350242}.
Nodes in this tree correspond to the tables of $\revc{\Tables{\jp}}$, and edges connecting the nodes represent equijoin conditions among them.
Assuming that queries are acyclic and contain no cross products,
one such tree always exists to capture $\revc{\JPreds{\jp}}$ precisely.
Indeed, there can be multiple alternative plan trees;
we defer the discussion of how to choose one to \Cref{sec:implementation}.
At a high level, we process $\revc{\queryset}[\revc{\jp}]$ using an in-order traversal of the plan tree starting from its root,
by calling its main workhorse, the recursive procedure $\ourrecursive$ (Line~\ref{alg:our_main:call_recurisve}),
to compute a count map for $\revc{\queryset}[\revc{\jp}]$.
This count map is then used to calculate the result counts for queries,
which can be done naively by summing up the counts at grid coordinates in each query hyperrectangle;
we describe an optimized implementation in \Cref{sec:implementation}.

\begin{figure}[t]
    \centering
    \includegraphics[width=\columnwidth]{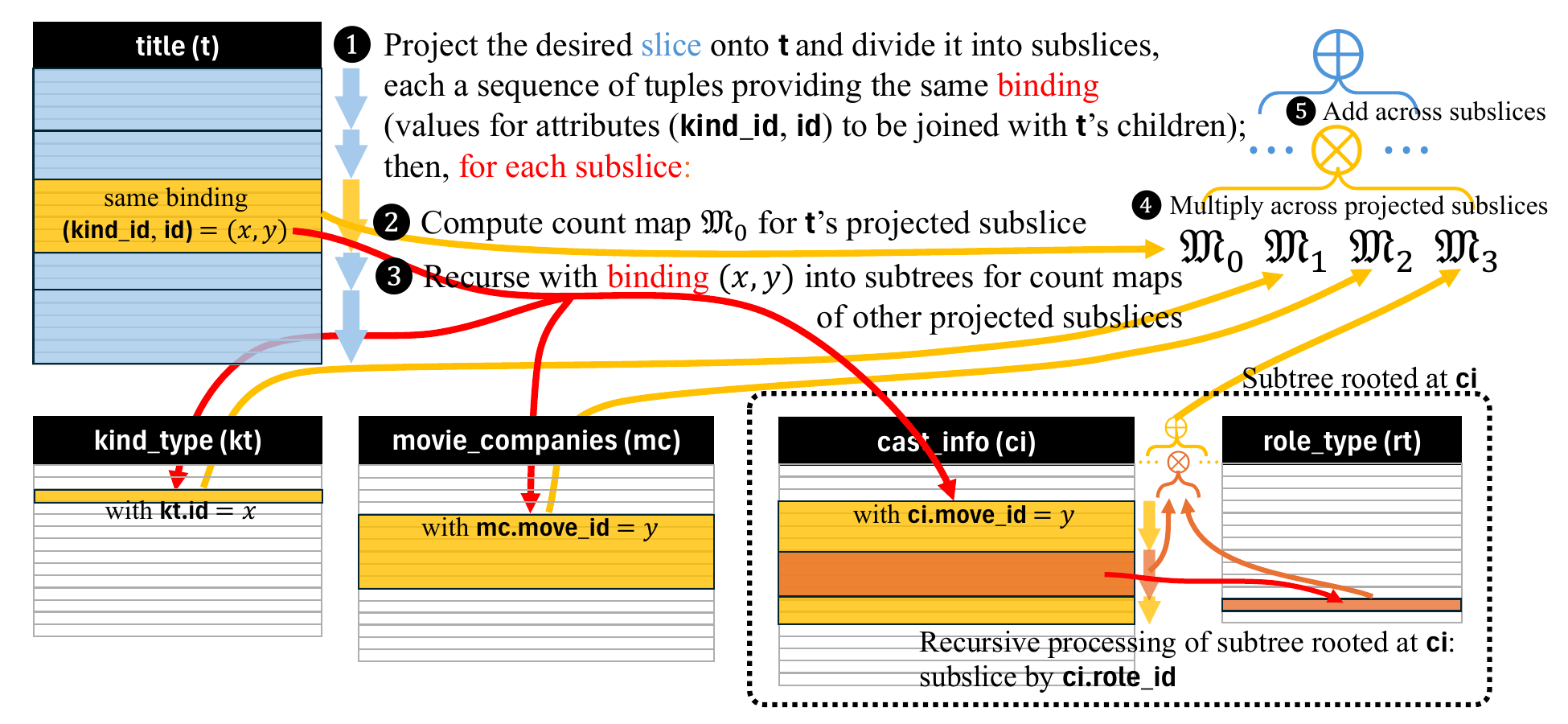}
    \vspace*{-4ex}
    \caption{\mdseries \common{Illustration of \Cref{alg:our_recursive}, continuing \Cref{fig:join_pattern_tree}.}}
    \label{fig:our_recursive}
    \vspace*{-2ex}
\end{figure}

\begin{algorithm}[t]
\caption{$\ourrecursive(\tab, u)$}\label{alg:our_recursive}
\begin{algorithmic}[1]
\Require{Mapping $u$ binds attributes a subset of $\tab$'s attributes to specific values.
    Implicitly, the function also has access to $\database$, $\revc{\jp}$ and its plan tree, and the quantization scales $\QuantScales_\revc{\jp}$.}
\Ensure{A count map $\CountMap$ over $\revc{\subtree}(\tab)$,
    complete w.r.t.\ result tuples of $\revc{\queryset[\jp]}$ restricted to $\revc{\subtree}(\tab)$ and consistent with $u$.}
\State $\CountMap \gets \emptyset$; \Comment{missing entries in all maps default to count $0$}
\For{each subsequence $\mathbf{S}[v]$ of entries $\langle v, \cdot, \cdot \rangle$ with the same $v$,
    returned by $\ProcessTable\big(\tab, u, \revc{\JAttrs{\jp}[\tab\,|\,\children(\tab)]}\big)$}\label{alg:our_recursive:loop}
    \State $\CountMap_0 \gets \{ \myvec{b} \mapsto c \mid \langle v, \myvec{b}, c \rangle \in \mathbf{S}[v] \}$;
    \For{each table $\tab_i \in \children(\tab)$}\label{alg:our_recursive:recursive_loop}
        \State $u_i \gets \left\{ A' \mapsto v(A) \;\middle|\;
            {\footnotesize
            \begin{aligned}
                & A \in \revc{\JAttrs{\jp}[\tab | \tab_i]} \land A' \in \revc{\JAttrs{\jp}[\tab_i | \tab]}\\
                & \land \revc{\JPreds{\jp}} \Rightarrow (A = A')
            \end{aligned}}
            \right\}$;\label{alg:our_recursive:project_binding}
        \State $\CountMap_i \gets \ourrecursive(\tab_i, u_i)$;
        \State $\CountMap_0 \gets \CountMap_0 \otimes \CountMap_i$;\label{alg:our_recursive:multiply}
    \EndFor\label{alg:our_recursive:recursive_loop_end}
    \State $\CountMap \gets \CountMap \oplus \CountMap_0$;\label{alg:our_recursive:add}
\EndFor\label{alg:our_recursive:loop-end}
\State \Return $\CountMap$;
\end{algorithmic}
\end{algorithm}
\ourrecursive\ (\Cref{alg:our_recursive}, illustrated in \Cref{fig:our_recursive})
is called on a plan tree node (table) $\tab$ and a filter specified by a mapping $u$ that binds some attributes of $\tab$ to specific values,
which defines a ``slice,'' or the subset $T$ of result tuples, of the join of $\revc{\subtree}(\tab)$ (set of tables in the subtree rooted at $\tab$) satisfying $u$.
The goal of \ourrecursive\ is to compute a count map for this slice $T$
(the call to $\revc{\treeroot}(\jp)$ has $u = \emptyset$, so its slice in fact contains all result tuples in the full join of $\revc{\Tables{\jp}}$).
As discussed in \Cref{sec:ours:count-maps}, processing the entirety of $T$ efficiently is hard;
instead, \ourrecursive\ partitions the slice into ``subslices,''
where each subslice $T_v$ binds a particular combination $v$ of values for $\tab$'s attributes that join with its children, i.e., $\revc{\JAttrs{\jp}[\tab\,|\,\children(\tab)]}$.
To this end, \ourrecursive\ calls \ProcessTable\ (\Cref{sql:count-table}) to enumerate all possible such bindings (Line~\ref{alg:our_recursive:loop}).
Each iteration of the loop (Lines~\ref{alg:our_recursive:loop}--\ref{alg:our_recursive:loop-end})
computes a count map for the subslice $T_v$ defined by a particular $v$.
By \Cref{lemma:mutiply-count}, this count map can be computed by projecting $T_v$ into ``projected sublices,''
one for $\tab$ and each of $\tab$'s child subqueries,
and multiplying the count maps of these projected subslices.
Conveniently, the call to \ProcessTable\ computes, for each projected subslice of $\tab$, the coordinate-count pairs to populate the count map.
We recursively call \ourrecursive\ on each child of $\tab$ to compute the count map for its projected subslice.
Line~\ref{alg:our_recursive:multiply}, using $\otimes$, combines these count maps
into a single count map for $T_v$.
Finally, applying \Cref{lemma:add-count}, Line~\ref{alg:our_recursive:add} uses $\oplus$ to accumulate the count maps for all $T_v$'s
into the final count map for $T$ to be returned.

\begin{example}
    Following \Cref{fig:our_recursive}, at the root \sql{title}, \ours{} iterates over all possible bindings for $(\sql{t.kind\_id}, \sql{t.id})$ in turn.
    Suppose the blue subslice fixes $\sql{t.kind\_id}=1, \sql{t.id}=1$.
    For \common{the subslice projected to \sql{t}}, suppose $\CountMap_{0} = \CountMap{\sql{t}} = \{(9) \mapsto 2, \dots\}$.
    We recursively call each child with the corresponding binding, as indicated by the arrows pointing to each child.
    For the ``projected subslices'' on children, we further assume that $\CountMap_{2} = \CountMap_{\sql{mc}}$ is the one shown in \Cref{fig:quantization}.
    \common{From the geometric view, $\CountMap_{2}$ is computed by the flattened hyperrectangle (in \Cref{fig:quantization}) with $\sql{mc.movie\_id} = \sql{t.id} = 1$.
    For illustration, let us say that $\CountMap_{1}$ ($\CountMap_{\sql{kt}}$) $=\{\dots, (1) \mapsto 1, \dots\}$ and $\CountMap_{3}$ ($\CountMap_{\sql{ci} \join \sql{rt}}$) $=\{\dots, (3, 3, 3) \mapsto 3, \dots\}$.
    }

    \common{This subslice contributes to the global count map $\CountMap$ as follows: 
    $\CountMap \gets \CountMap \oplus (\CountMap_0 \otimes \dots \otimes \CountMap_3)$.
    We zoom in on one example---how the global count map entry at grid coordinate $(9, 1, 2, 2, 3, 3, 3)$ is updated.
    Concretely, this global grid coordinate is the concatenation of
    $(9)$ from $\CountMap_{0}$, $(1)$ from $\CountMap_{1}$, $(2, 2)$ from $\CountMap_{2}$, and $(3, 3, 3)$ from $\CountMap_{3}$.
    Recall that the counts in $\CountMap_{0}, \ldots, \CountMap_{3}$ at these coordinates are $2, 1, 2, 3$, respectively;
    therefore, the contribution to the global count map entry is $2 \times 1 \times 2 \times 3 = 12$.}
    After processing this subslice, \ourrecursive{} proceeds to the next $(\sql{t.kind\_id}, \sql{t.id})$ combination in order.
\end{example}

\begin{sqllisting}[float=t]{\mdseries SQL code for $\ProcessTable(\tab, u, \revc{\revc{\attrs}^{\texttt{out}}})$.}{sql:count-table}`
`\textnormal{\textbf{Input:} Mapping $u$ binds attributes $\Jin_1, \Jin_2, \ldots$ to specific values, and $\revc{\revc{\attrs}^{\texttt{out}}} = \{\Jout_1, \Jout_2, \ldots\}$ specifies the set of attributes for partitioning result entries. The quantization scales are $\QuantScales_\revc{\jp}[\tab] = \{ \QuantScale_{S_1}, \QuantScale_{S_2}, \ldots \}$.}`
`\textnormal{\textbf{Output:} Result entries have the form $\langle v, \myvec{b}, c\rangle$, sorted by $v$, where $v$ is a mapping from $\revc{\revc{\attrs}^{\texttt{out}}}$ to values, $\myvec{b}$ is a grid coordinate for $\QuantScales_\revc{\jp}[\tab]$, and $c$ is the associated count.}`
`\noindent\rule{0.98\columnwidth}{0.2pt}`
SELECT $\Jout_1$, $\Jout_2$, ..., B1, B2, ..., COUNT(*)
FROM (
    SELECT $\Jout_1$, $\Jout_2$, ...,
        quantize($S_1, \QuantScale_{S_1}$) AS B1, quantize($S_2, \QuantScale_{S_2}$) AS B2, ...
    FROM $\tab$
    WHERE $\Jin_1$ = $u(\Jin_1)$ AND $\Jin_2$ = $u(\Jin_2)$ AND ...
) AS TMP
GROUP BY $\Jout_1$, $\Jout_2$, ..., B1, B2, ...
ORDER BY $\Jout_1$, $\Jout_2$, ...;
\end{sqllisting}

Delving into \ProcessTable\ (\Cref{sql:count-table}),
we see how a single SQL query over a table $\tab$ performs ``subslicing'' (by $\revc{\revc{\attrs}^\texttt{out}}$) and computes the count maps of all projected subslices.
In the \sql{FROM} subquery that defines \sql{TMP},
the \sql{WHERE} condition applies the attribute binding $u$ that defines $T$, the slice of interest;
the \sql{SELECT} clause uses a user-defined function \sql{quantize} to check each selection attribute value against a quantization scale and returns its grid coordinate.
The outer query groups the \sql{TMP} tuples by $\revc{\revc{\attrs}^\texttt{out}}$ to perform subslicing,
and then by their grid coordinates to compute tuple count per grid cell for each projected subslice.
Finally, \sql{ORDER} \sql{BY} ensures that entries for the same projected subslice,
i.e., those with the same binding $v$ for $\revc{\revc{\attrs}^\texttt{out}}$,
are consecutive in the output.
This ordering allows the caller \ourrecursive\ to detect where a sequence of entries with a new $v$ starts,
so it can start processing a new subslice.

\section{Additional Optimization and Implementation Details}
\label{sec:implementation}

This section fills in some of the details about \ours\ not covered by \Cref{sec:ours}
and describes additional optimizations needed to make \ours\ practical for query workloads for training learned CE models.
Before delving into details, we note that \ours\ takes the high-level approach of leveraging DBMS for processing,
not only because it houses the data, but also because it already provides versatile support for indexing and querying.
To make \ours\ easy to adopt, we do not modify any DBMS internals
but instead implement, in a client application, parts of the algorithm that are inefficient for the DBMS.
Many optimizations thus involve balancing database- and client-side processing options and mitigating the overhead of interfacing them.
\revc{Overall, we perform \ProcessTable\ (\Cref{sql:count-table}) inside the DBMS, using cursors to fetched join values, quantization grid coordinates, and associated counts.
The remainder of \ourrecursive\ (i.e., merging of count maps along the tree structure)
and the remainder of \ours\ (i.e., \ComputeCounts) are performed on the client side.
The associated optimizations are presented below.}
While \ours\ is currently implemented on top of PostgreSQL, the ideas in this section generalize to other DBMSs,
although details may differ.

\revc{Note that \ours\ is not optimized for join patterns characterized by
queries exhibiting little opportunity for shared processing
(e.g., when there are too few of them, or their predicates overlap very little),
or an underlying full join that is small and cheap to compute to begin with.
In such cases,
\ours\ may not outperform baselines \sequentialproc\ or \postfilter.
Therefore, we also have developed a method called \hybrid,
which acts as an optimizer to choose the appropriate method to use among \sequentialproc, \postfilter, and \ours\ for a given join pattern.
In our experimental comparison of these methods in \Cref{sec:experiment},
we empirically analyze the cases where \ours\ is not optimal (\Cref{subsec:exp_per_join_pattern}):
they follow our characterizations above and play only a small role in the overall performance of query workloads found in the learned CE literature.
Furthermore, despite \hybrid's ability to make on per-pattern decisions, it offers no consistent improvement over \ours\ overall.
This observation speaks to the effectiveness of \ours, even without \hybrid, for handling practical CE training workloads.}
\common{More details on \hybrid\ can be found in the full version~\cite{fullversion}.}

\subsection{Choosing Execution Plan for a Join Pattern}
\label{sec:implementation:plan}

Give a join pattern $\revc{\jp}$, the cost of \ourrecursive\ to process $\revc{\queryset[\jp]}$ depends on the choice of a tree-based execution plan, introduced in \Cref{sec:ours:algo}.
A cost-based decision requires fine-grained knowledge of the data distribution, which itself is expensive to acquire.
Instead, we follow some heuristics below to construct a tree-based plan.

The first heuristic is to pick tree root to be the table $\tab$ with the highest degree in $\revc{\jp}$'s join graph:
i.e., the root joins with the largest number of other tables in $\revc{\Tables{\jp}}$ according to $\revc{\JPreds{\jp}}$.
The rationale is to maximize the saving opportunities identified by \Cref{lemma:mutiply-count}:
a root with a large in-degree can potentially avoid enumerating more cross products.
\common{
Geometrically, this heuristic ensures that, during recursion, the grid coordinates in each descendant lie in low-dimensional subspaces.
}
Once we pick the root, the rest of the tree structure follows naturally from $\revc{\JPreds{\jp}}$.
Since we assume queries to be acyclic, no additional join conditions are needed beyond those corresponding to the tree edges.
For example, in \Cref{fig:join_pattern_tree},
even though \sql{mc} and \sql{ci} can join on \sql{mc.movie\_id = ci.movie\_id},
there is no edge between them in the tree,
because the two join conditions from root \sql{t} to \sql{mc} and \sql{ci}, involving \sql{t.id}, together imply \sql{mc.movie\_id = ci.movie\_id}.

The next optimization then marks certain tree edges that correspond to joining a foreign key (FK) of the parent table $\tab$ to a primary key (PK) of the child table $\tab'$.
For such a marked FK-PK edge, when \ourrecursive\ processes $\tab$, it will simultaneously process $\tab'$:
on Line~\ref{alg:our_recursive:loop}, \ProcessTable\ will additionally query $\tab'$ and compute its count map along with that for $\tab$
(with appropriate changes to \Cref{sql:count-table});
and on Line~\ref{alg:our_recursive:recursive_loop}, children of $\tab'$ will be directly included, instead of $\tab'$ itself.
The justification behind this optimization is that,
because of the FK-PK join, each $v$ would yield only one joining tuple from $\tab'$, meaning there is no cross product to avoid in the first place.
Therefore, we should ``short-cut'' this edge to eliminate the associated overhead.%
\footnote{Note that the direction is important:
    a PK-FK edge, where the parent table has the primary key, would be a bad candidate to ``short-cut,'' with an opposite argument.}
For simplicity, our implementation currently restricts this optimization to edges incident to the root,
but applying it to the entire tree is possible as future work.

The last heuristic dynamically reorders subtrees under each node.
We start with an arbitrary ordering.
For a node $\tab$, after some iterations of the loop on Lines~\ref{alg:our_recursive:loop}--\ref{alg:our_recursive:loop-end} of \Cref{alg:our_recursive},
we count, for each child $\tab_i$, how many settings of $v$ yield no tuples from $\tab_i$.
Then, we reorder the children so that those with more such cases come first.
Intuitively, they are more likely to produce an empty count map, with which we can ``short-circuit'' the sequence of multiplies.

\subsection{Mitigating SQL Querying Overhead}
\label{sec:implementation:sql}

\paragraph{Server- vs.\ Client-Side Cursors}

When \ours\ runs a SQL query on the underlying database, we can use either a server-side cursor~\cite{server_side_cursors} or a client-side one~\cite{client_side_cursors}.
A client-side cursor fetches the entire query result set into client memory.
A server-side cursor, in contrast, allows the client to iterate row-by-row without materializing a potentially massive result set.
While server-side cursors offer more flexibility and improve scalability,
they are considerably slower and require state to be maintained on the database server;
a large number of concurrent server-side cursors can cause performance issues.
Therefore, \ours\ uses a server-side cursor for \ProcessTable\ on the root,
but switches to client-side cursors for all other calls.
The rationale is that
the call on the root table (from Line~\ref{alg:our_main:call_recurisve} of \Cref{alg:our_main}) has no bound attributes ($u = \emptyset$)
and may generate a large result set.
In contrast, other calls access much fewer tuples per subslice,
and even with the batching optimization below, their result sizes remain manageable for client-side cursors.

\paragraph{Batching of \ProcessTable}

During recursive execution of \ourrecursive\ over the plan tree,
the overhead of issuing many SQL queries via \ProcessTable\ can accumulate,
especially when many of these queries, with specific attribute bindings, return small result sets.
To mitigate this overhead, we batch multiple invocations of \ProcessTable\ for the same table $\tab$,
using a set of bindings $U$ rather than a single binding $u$, into one SQL query.
Specifying the combined condition precisely in the \sql{WHERE} clause of \Cref{sql:count-table} is generally infeasible,
as it would have to enumerate all bindings in $U$.
Instead, we compute the minimum and maximum elements%
\footnote{If each binding involves multiple attributes, we order the bindings lexicographically.}
of $U$ and use $[\min(U), \max(U)]$ as a safe but possibly imprecise range bound in \sql{WHERE},
i.e., \sql{WHERE} $(\Jin_1, \Jin_2, \ldots)$ \sql{BETWEEN} $\min(U)$ \sql{AND} $\max(U)$.
We extend the query to additionally group by $\Jin_1, \Jin_2, \ldots$ and return them,
such that \ProcessTable\ can, using these attribute values, filter out result tuples that do not belong to $U$.

In \ourrecursive, we implement this optimization by collecting $\beta$ consecutive settings of $v$ into a set $V$ for processing as a batch,
analogous to a $\beta$-fold loop unrolling on Lines~\ref{alg:our_recursive:loop}--\ref{alg:our_recursive:loop-end} of \Cref{alg:our_recursive}.
For each child $\tab_i$, we project $V$
to obtain the set $U_i$ of unique bindings on its join attributes with $\tab$ (Line~\ref{alg:our_recursive:project_binding}),
and invoke \ProcessTable\ on $\tab_i$ with $U_i$.
The detailed algorithm is presented \reva{in the full version}~\cite{fullversion}.

We now provide further analysis of this batching optimization.
First, beyond reducing the overhead of issuing many queries,
batching promotes reuse of \ProcessTable\ calls on the same table and attribute binding.
Without batching, consecutive settings of $v$ in the loop on Lines~\ref{alg:our_recursive:loop}--\ref{alg:our_recursive:loop-end} of \Cref{alg:our_recursive}:
it is likely that they agree on most components since they are sorted.
Therefore, when $v$ is projected to obtain some $u_i$, the same $u_i$ setting in the previous iteration may be used again to call \ProcessTable\ on the same table $\tab_i$, leading to waste unless we cache the results from previous calls.
With batching, however, such reuse occurs naturally.
So we have not implemented additional caching mechanisms for \ours,
although further exploration will be a good future work.

Second, compressing a set of bindings into a single range may introduce false positives.
Indeed, $\tab_i$ may always contain some tuples whose join attribute values lie in $[\min(U_i), \max(U_i)]$
but are not contained in $U_i$, because $U_i$ is generated from $V$ and subject to all other constraints on the slice.
Nonetheless, the concern with false positives is at least partly alleviated by the fact that \ourrecursive\ processes settings of $v$ in order,
meaning that it will call \ProcessTable\ on $\tab_i$ with ``clustered'' settings of $u$ in $U$.

\paragraph{Implementation of \sql{quantize}}

Since the UDF \sql{quantize} in \Cref{sql:count-table} is invoked many times, efficient implementation is crucial for performance.
We implement \sql{quantize} as C-language UDFs in PostgreSQL~\cite{postgresql_c_function}, with one variant for each SQL data type.
The quantization scale is implemented as an array of sorted boundary points, and a binary search is used to determine the correct bucket id.
These C-language UDFs run as compiled code natively inside the server process, greatly outperforming SQL-based UDFs.
\ours\ creates these extensions at the beginning of its execution and drops them at the end.
\reva{More generally, \sql{quantize} can be implemented in other DBMSs using available extensions, including CLR (common language runtime) integration for Microsoft SQL Server~\cite{sql_server_clr} or external procedures for Oracle Database~\cite{oracle_external_procedures}.}

\subsection{Implementation of \ComputeCounts}
\label{sec:implementation:ComputeCounts}

\ComputeCounts\ is invoked by \ours\ to compute the result counts of queries in $\revc{\queryset[\jp]}$ given the final count map $\CountMap$ computed by \ourrecursive.
Conceptually, it performs a ``join'' between a set of query hyperrectangles ($\revc{\queryset[\jp]}$) and a set of weighted points ($\CountMap$), and reports the total weights within each hyperrectangle.
Many processing strategies are possible, and the best depends on query and data distributions.
For our target of query workloads for training learned CE models, we observe that $|\revc{\queryset[\jp]}|$ is moderate (in hundreds or thousands) and $\CountMap$ tends to be sparse (32\% on average across join patterns in JOB's \texttt{synthetic}).
After experimenting with several methods,
we have found a simple method based on nested loops to be the most effective for this setting.
Basically, for each $\query \in \revc{\queryset[\jp]}$ and for each non-zero entry $\myvec{b} \mapsto c$ of $\CountMap$,
we check whether, for every dimension of $\query$'s hyperrectangle with a range bound,
the corresponding coordinate in $\myvec{b}$ falls within this bound.
If yes, we add the entry's count $c$ to $\query$'s running total.
While \ours\ is implemented in Python, we specifically use Numba's just-in-time compilation~\cite{numba} for \ComputeCounts.
With the help from Numba's compiler optimizations,
this simple implementation was able to beat more sophisticated methods.

\section{Experiments}
\label{sec:experiment}

We compare \sequentialproc{},
\revb{\postfilter,\footnote{\revb{
    We have also evaluated an alternative implementation of \postfilter{} based on materialized views.
    The details and results, reported in the full version~\cite{fullversion}, show that materialization introduces additional storage costs and preparation-time overhead,
    without providing overall performance benefits.
    Hence, \Cref{sec:experiment} focuses on the current implementation of \postfilter{}.
}}}
and \ours{}.
All algorithms are implemented in Python3,
using \texttt{psycopg2}~\cite{psycopg2} to connect to PostgreSQL V16.8 with default configuration. 
All experiments are conducted on a Linux server with an Intel(R) Xeon(R) Gold 5215 CPU (40 logical cores, 2.50GHz), \revb{256GB of main memory}, and 1TB of disk storage.
We set the batching parameter $\batchparam{}$ (\Cref{sec:implementation:sql}) to 50,000 for \ours{} for all experiments and present results for different $\batchparam$'s in \reva{the full version}~\cite{fullversion}.
Code is available \reva{in the GitHub repository}~\cite{bacon_repo}.

We test three popular datasets: IMDB~\cite{10.14778/2850583.2850594}, STATS~\cite{10.14778/3503585.3503586}, and DSB~\cite{10.14778/3484224.3484234} (of scaling factor 2).
They are loaded and initialized using \reva{the provided scripts}~\cite{imdb_loading_script, stats_loading_script, dsb_loading_script}, and the algorithms use exactly the indices provided therein without creating any additional ones.

We evaluate the algorithms on nine publicly available workloads in \Cref{tab:query_workloads}. 
These workloads span a wide range of sizes, query templates, join structures, predicate types, and result sizes, and are widely used for training and evaluating cardinality estimation techniques~\cite{10.14778/3503585.3503586}.
Among the workloads, only \texttt{dsb-grasp-20k} is post-processed\reva{~\cite{script_for_dsb_grasp_20k} from the source file~\cite{original_grasp_sql}}. 
Specifically, CE work, SeConCDF~\cite{10.14778/3725688.3725708} and GRASP~\cite{10.14778/3742728.3742745}, have numerical/categorical definitions for several attributes that are originally defined as \sql{char(...)} in DSB (e.g., \sql{cd.cd\_education\_status}).
As these transformations are not clearly specified, we remove the queries involving such attributes, and then randomly sample 20k queries to ensure that baselines complete within a reasonable time.
We additionally construct three new workloads based on \texttt{job-light}, used exclusively in \Cref{subsec:exp_scalability}.

Running time is our main performance metric.
For \sequentialproc, end-to-end running time is the sum over all queries.
For \postfilter\ and \ours, it includes pre-processing and time for each join pattern.
To prevent baselines from running for too long, we cap \sequentialproc\ at 900s per query, and \postfilter\ at 3{,}600s per join pattern;
therefore, some baseline numbers are only lower bounds.

\revb{In terms of client-side memory usage,
because \ours{} processes each join pattern separately,
memory peaks at the single most complex join pattern in the workload.
Across all nine workloads, the maximum memory footprint of \ours{} occurs for a join pattern in \texttt{synthetic}, at 1.7GB.
We do not report memory usage for \postfilter\ and \sequentialproc, as they executed entirely inside the database server.}

\paragraph{\revc{Single-Threaded vs.\ Parallel Execution}}
\revc{
Results below focus on single-threaded execution to isolate the algorithmic differences of \ours{} compared with the baselines,
without the potentially confounding factor of parallelization.
We disable parallelism in PostgreSQL by setting \sql{max\_parallel\_workers\_per\_gather} to 0,
and restrict all client-side code to a single thread.
Besides this single-threaded setting, we have also evaluated all methods in a parallel setting.
Under the parallel setting, \sequentialproc{} and \postfilter{} fully leverage PostgreSQL's built-in parallelism;
for \ours, we use a simple multi-process design in which each process handles a subset of join patterns
(more sophisticated parallelization strategies for \ours{} are possible as future work).
Briefly, the results show that, even with this simple strategy,
\ours\ achieves end-to-end speedups of 2$\times$ to 598$\times$ compared with \sequentialproc{},
consistent with the single-threaded results shown below in \Cref{subsec:exp_end_to_end}.
Additional details on the parallel setting can be found in the full version~\cite{fullversion}.}

\subsection{End-to-End Running Time}
\label{subsec:exp_end_to_end}

\Cref{tab:end_to_end_runtime} reports the end-to-end running times of all three approaches across all nine real workloads.
Lower-bound values (followed by ``+'') are reported if a baseline runs out of time.
We additionally show the relative speedups of \ours{} over \sequentialproc{}, which range from approximately 2$\times$ to 178$\times$.
Speedups over \postfilter{} are omitted, as \sequentialproc{} outperforms \postfilter{} on most long-running workloads.
Overall, \ours{} consistently outperforms both baselines and never times out:
the worst per-join-pattern overhead of \ours{} across all workloads is below 380s, far below the thresholds for baselines.
The observed performance trends between \sequentialproc{} and \postfilter{} align with our analysis in \Cref{sec:prelim}.
In particular, \postfilter{} is advantageous for join patterns with many queries---where predicates tend to cover larger regions---and without intermediate join blow-ups (e.g., \texttt{synthetic} and \texttt{dsb-grasp-20k}).
In the remaining cases, \sequentialproc{} performs substantially better due to selection push-down in each query.
\ours{} is able to effectively combine the strengths of both \sequentialproc{} and \postfilter{}.

\begin{table}[t]
    \caption{\mdseries End-to-end running times (seconds, rounded) and relative speedups. +: lower-bound values, as \sequentialproc{} is capped at 900s per query and \postfilter{} at 3{,}600s per join pattern.}
    \label{tab:end_to_end_runtime}
    \vspace{-1em}
    \small
    \begin{tabular}{c||r|r|r|r|r|}
          Query Workload & \sequentialproc{} & \postfilter{} & \ours{} & \begin{tabular}{@{}r@{}}Speedup \\vs. \sequentialproc{}\end{tabular}
           \\\hline\hline
            \texttt{synthetic} & 19,663 & 11,329+ & 1,574 & $\uparrow$ 12.49 $\times$
            \\\hline
            \texttt{scale} & 21,100+ & 26,342+ & 3,475 &  $\uparrow$ 6.07+ $\times$
            \\\hline
            \texttt{job-light} & 1,620+ & 13,662+ & 728 & $\uparrow$ 2.22+ $\times$
            \\\hline
            \texttt{job-light-single} & 249 & 13 & 11 & $\uparrow$ 22.50 $\times$
            \\\hline
            \texttt{job-light-join} & 8,601+ & 18,353+ & 1,541 &$\uparrow$ 5.58+ $\times$
            \\\hline
            \texttt{stats-ceb} & 4,274+ & 47,353+ & 58 & $\uparrow$ 73.62+ $\times$
            \\\hline
            \texttt{stats-ceb-single} & 18 & 2 & 2 & $\uparrow$ 8.70 $\times$
            \\\hline
            \texttt{stats-ceb-join} & 35,292+ & 53,363+ & 198 & $\uparrow$ 178.14+ $\times$
            \\\hline
            \texttt{dsb-grasp-20k} & 19,746 & 4,211 & 928 & $\uparrow$ 21.28 $\times$
         \\\hline
    \end{tabular}
\end{table}

Since both \postfilter{} and \ours{} are designed around the notion of join patterns, we further analyze the results on a per-join-pattern basis in \Cref{subsec:exp_per_join_pattern}.
Before doing so, we briefly summarize the other overheads---which are negligible compared with per-join-pattern running time---and omit them from ensuing discussions.
\sequentialproc{} incurs no such overheads.
Both \postfilter{} and \ours{} require loading and parsing SQL queries,  extracting join patterns, and grouping queries accordingly, with complexity linear in the number of queries, tables, and selection attributes.
This pre-processing step happens at the beginning when processing each join pattern, and typically completes within a few seconds; the only exception is \texttt{dsb-grasp-20k}, where it reaches $\approx$20s due to the large number of queries (but is still small relative to the end-to-end running time).

\subsection{Per-Join-Pattern Result}
\label{subsec:exp_per_join_pattern}

In this section, we analyze the results on a per-join-pattern basis.
Although \sequentialproc{} does not operate on join patterns, we group its results accordingly to facilitate direct comparison with the others.

\begin{table}[t]
    \caption{\mdseries Number of timed-out cases. In ($a$, $b$), $a$ is the number of timed-out queries, and $b$ is the number of join patterns that contain any timed-out queries.}
    \label{tab:unavailable_results}
    \vspace{-1em}
    \small
    \begin{tabular}{c||r|r|r|r|}
          Query Workload & \# queries & \begin{tabular}{@{}r@{}}\# join\\ patterns\end{tabular} & \sequentialproc{}  & \postfilter{} 
           \\\hline\hline
            \texttt{synthetic} & 5,000 & 16 & (0, 0) & (251, 1)
            \\\hline
            \texttt{scale} & 500 & 31 & (20, 7) & (84, 6) 
            \\\hline
            \texttt{job-light} & 70 & 18 & (1, 1) & (8, 2)
            \\\hline
            \texttt{job-light-single} & 254 & 1 & (0, 0) & (0, 0) 
            \\\hline
            \texttt{job-light-join} & 696 & 27 & (3, 1) & (32, 3)
            \\\hline
            \texttt{stats-ceb} & 146 & 58 & (3, 2) & (30, 11) 
            \\\hline
            \texttt{stats-ceb-single} & 632 & 1 & (0, 0) & (0, 0) 
            \\\hline
            \texttt{stats-ceb-join} & 2,603 & 120 & (22, 7) & (166, 13) 
            \\\hline
            \texttt{dsb-grasp-20k} & 20,000 & 16 & (0, 0) & (0, 0) 
         \\\hline
    \end{tabular}
\end{table}

We first report the number of timed-out cases in \Cref{tab:unavailable_results}.
\ours{} never times out.
Overall, \postfilter{} times out more frequently: since it always materializes the unfiltered join, if that times out, all queries associated with the pattern time out.
Additionally, workloads on STATS exhibit more time-outs, primarily due to larger intermediate joins than those on IMDB, as noted \reva{by~\citet{10.14778/3503585.3503586}}.

\begin{figure*}[t]
    \centering
    \includegraphics[width=\textwidth]{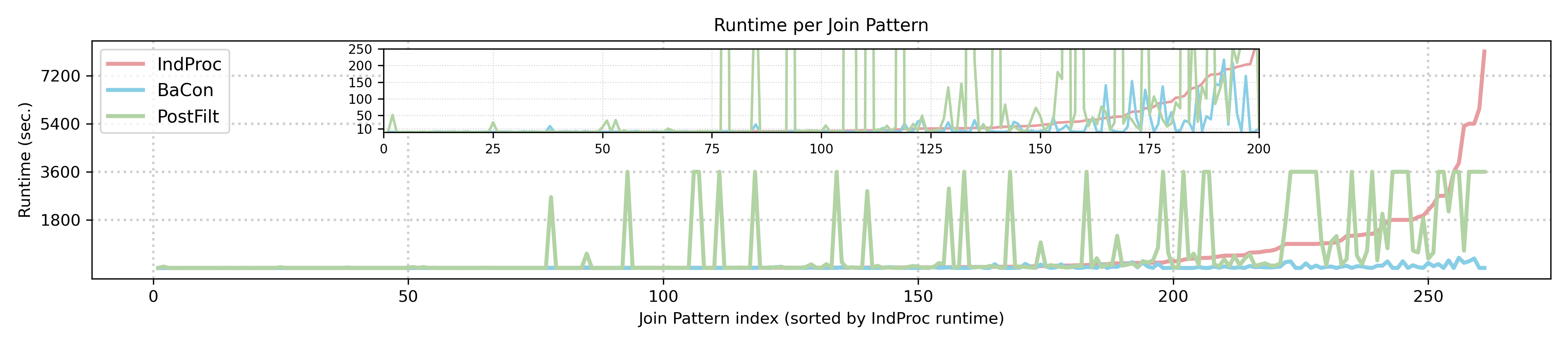}
    \vspace*{-2ex}
    \caption{\mdseries Running time (seconds) per join pattern across 283 join patterns from 9 workloads. Join patterns are ordered by \sequentialproc's times. The inset is the zoom-in of the first 200 indices.}
    \label{fig:ordered_runtime}
\end{figure*}

Next, we analyze the per-join-pattern overhead across all workloads.
For clarity, consistent with the pre-defined time-out thresholds, we assign 3,600s to timed-out join patterns in \postfilter{} and 900s to timed-out queries in \sequentialproc{}; actual running times will be higher.
\Cref{fig:ordered_runtime} reports the running times of all join patterns, ordered by \sequentialproc's times.
Before index 120, all join patterns under \sequentialproc{} complete within 10s; beyond this point, running times increase markedly.
The inset zooms in the region before index 200, where \ours{} closely tracks \sequentialproc{}, sometimes faster and sometimes slower within a bounded margin, while \postfilter{} already times out on several patterns.
Between indices 160 and 200, \ours{} exhibits a small number of clear regressions relative to \sequentialproc{}, with the largest gap (98s) at index 165;
we analyze these cases in detail later.
Outside the zoomed region, where join patterns run substantially longer, \ours{} consistently and significantly outperforms \sequentialproc{} and \postfilter{}.

\begin{figure*}[t]
    \centering
    \includegraphics[width=\textwidth]{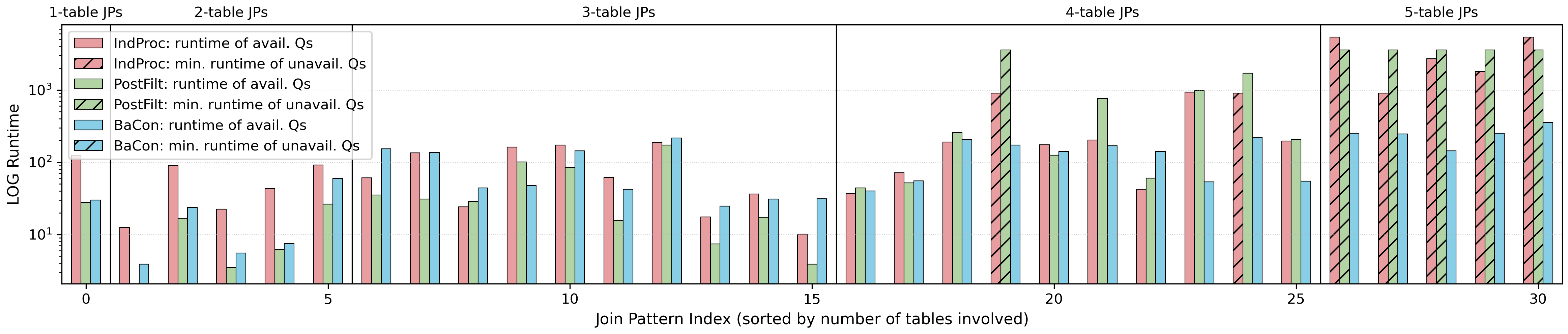}
    \vspace*{-3ex}
    \caption{\mdseries Log-scale running time per join pattern in \texttt{scale}. Join patterns are ordered by the number of tables involved, and the figure is partitioned accordingly (labels shown above). Hatched bars (baselines only) indicate the lower bounds for timed-out cases.}
    \label{fig:scale_per_jp}
\end{figure*}

Additionally, we present a detailed per-join-pattern analysis for \texttt{scale} and \texttt{job-light} here; results for the remaining workloads are in \reva{the full version~\cite{fullversion}}.
As shown in \Cref{fig:scale_per_jp}, \texttt{scale} contains join patterns involving 1 to 5 tables, with more tables generally run longer.
Timed-out queries arise mainly from patterns with many tables.
For most long-running join patterns, \ours{} consistently outperforms the baselines.
For fast patterns, \ours{} remains competitive, as the time gaps are small.
However, there are several patterns where \ours{} is noticeably slower than either baseline.
Taking join pattern 6 as an example, \sequentialproc{}, \postfilter{}, and \ours{} take 61s, 35s, and 153s, respectively.
This pattern has conditions \sql{ci.movie\_id = t.id AND mc.movie\_id = t.id} and only five queries.
A breakdown of \ours{}'s running time shows 118s (76\%) for executing SQLs, 5s (3\%) for aggregating rows by join attribute bindings and coordinates, and 18s (15\%) for merging count maps;
thus, SQL execution dominates and already exceeds the running times of the baselines.
This regression stems from two factors.
First, algorithmically, enumerating tuples from each table and quantizing them can be less efficient than \sequentialproc{} when predicates are selective and overlaps are limited.
Second, although \ours\ performs less work than \postfilter, \ours's cursor invocations and row fetching/parsing incur more overhead than \postfilter's server-side processing.

\begin{figure}[t]
    \centering
    \includegraphics[width=\columnwidth]{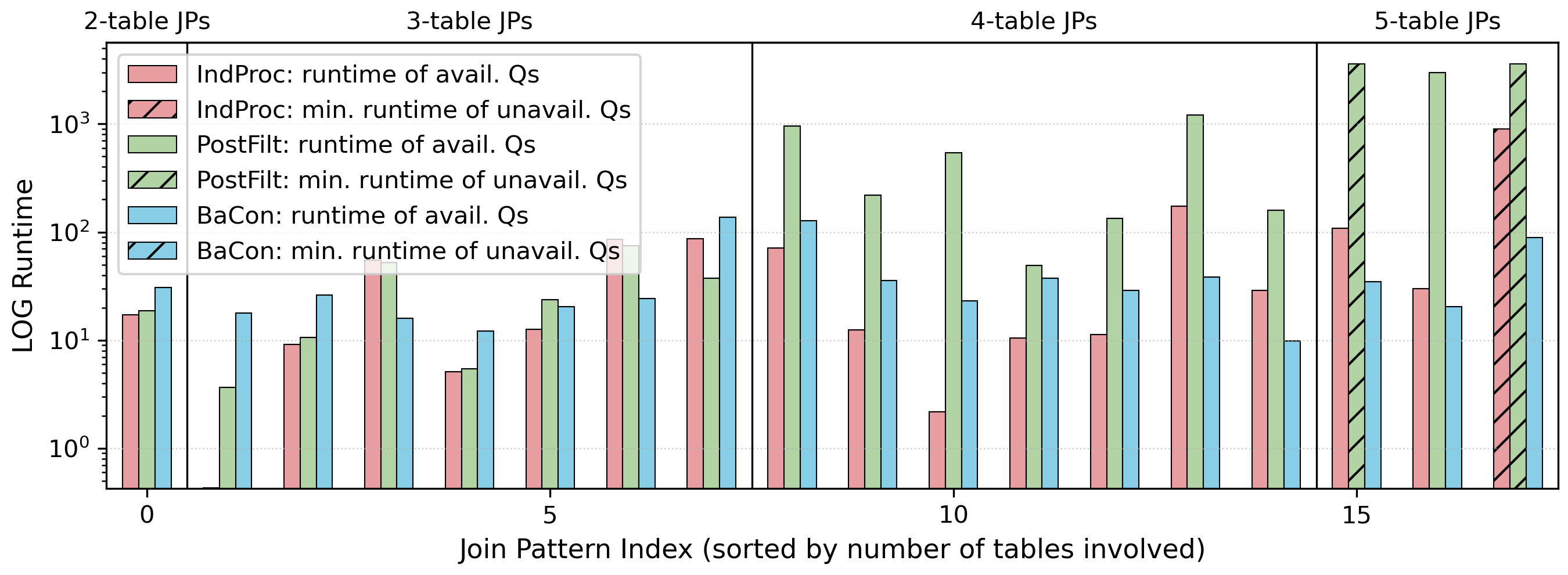}
    \vspace*{-3ex}
    \caption{\mdseries Log-scale running time per join pattern in \texttt{job-light}, with same format as \Cref{fig:scale_per_jp}.}
    \label{fig:job_light_per_jp}
\end{figure}

We have similar observations from \Cref{fig:job_light_per_jp}. 
As discussed \reva{by \citet{10.14778/3503585.3503586}}, \texttt{job-light} is a classical but relatively simple workload with 1 to 8 queries for each join pattern.
The true cardinality range is also a magnitude smaller than that of \texttt{stats-ceb}.
Therefore, \sequentialproc{} consistently performs well on all join patterns, except one severe time-out for join pattern 17.
However, \ours{} remains competitive across join patterns and never times out.
Regression compared with baselines is limited:
in the worst case, \ours{} is 55s slower compared with \sequentialproc{} on join pattern 8, and 100s compared with \postfilter{} on join pattern 7.

Overall, consistent with the results in \Cref{subsec:exp_end_to_end}, \ours{} does well across join patterns: it significantly outperforms the baselines on expensive patterns and remains acceptable on cheap ones.

\subsection{Scalability}
\label{subsec:exp_scalability}

In this section, we evaluate the scalability of the three approaches using three synthetic workloads, denoted \texttt{job-light-*}.
We choose \texttt{job-light} as the reference because, as shown in \Cref{fig:job_light_per_jp}, \sequentialproc{} performs consistently well, except for a single join pattern that times out. 
Specifically, on \texttt{jog-light}, \ours{} outperforms \sequentialproc{}{} in only 7 out of the 18 (39\%) join patterns.
Earlier, we have made the observation that the baselines---especially \sequentialproc{}---can outperform \ours{} on simple join patterns,
\revb{e.g., where the number of queries is small and intermediate join sizes remain moderate.
In particular, if the number of queries is small, there is less opportunity for sharing,
and \sequentialproc's simple approach of optimizing and evaluating queries one by one works fine.}
Here, we further investigate how each approach scales as the workload size increases while preserving the original distribution.

We construct workloads with 1,000, 2,000, and 4,000 queries each.
Queries are generated randomly and independently.
We randomly select a query from \texttt{job-light}, reuse its template (and thus its join pattern),
but modify its predicate constants:
1)~for a constant in equality and inequality predicates, we replace it with a random value from the active domain of the corresponding attribute;
2)~for range predicates, we randomize the left endpoint and adjust the right accordingly to preserve the range length.

\begin{table}[t]
    \caption{\mdseries 
    End-to-end running times (seconds, rounded) and relative speedups. +: lower-bound values, as \sequentialproc{} is capped at 900s per query and \postfilter{} at 3{,}600s per join pattern.}
    \label{tab:end_to_end_runtime_scalability}
    \vspace{-1em}
    \small
    \begin{tabular}{c||r|r|r|r|r|}
          Query Workload & \sequentialproc{} & \postfilter{} & \ours{} & \begin{tabular}{@{}r@{}}Speedup \\vs. \sequentialproc{}\end{tabular}
           \\\hline\hline
            \texttt{job-light-1k} & 6,208 & 22,490+ & 1,020 & $\uparrow$ 6.09 $\times$
            \\\hline
            \texttt{job-light-2k} & 13,818 & 26,923+ & 1,051 & $\uparrow$ 13.15 $\times$
            \\\hline
            \texttt{job-light-4k} & 23,617+ & 33,981+ & 1,109 & $\uparrow$ 21.29+ $\times$
         \\\hline
    \end{tabular}
\end{table}

\begin{table}[t]
    \caption{\mdseries End-to-end (E2E) running times for \ours{} and \hybrid{}, with the number of join patterns handled by each algorithm. \textsc{I/P/B} denotes \sequentialproc{}/\postfilter{}/\ours{}.}
    \label{tab:end_to_end_runtime_hybrid}
    \vspace{-1em}
    \small
    \begin{tabular}{c||r|r|r|}
          Query Workload & \begin{tabular}{@{}r@{}}\ours{} E2E\\running time\end{tabular} &  \begin{tabular}{@{}r@{}}\hybrid{} E2E\\running time\end{tabular} & \begin{tabular}{@{}c@{}}\# join patterns\\ using \textsc{I/P/B} \end{tabular}
           \\\hline\hline
            \texttt{scale} & 3,475 & 3,319 & 0/9/22
            \\\hline
            \texttt{stats-ceb} & 58 & 195 & 3/1/54
            \\\hline
            other 5 & - & - & 0/0/all
         \\\hline
    \end{tabular}
\end{table}

\Cref{tab:end_to_end_runtime_scalability} shows end-to-end running times and relative speedups. 
As expected, \sequentialproc{}'s running time grows roughly linearly with workload size, since queries are processed independently.
\postfilter{} grows more slowly because its dominant cost---computing the full join---is amortized across queries with the same join pattern; nevertheless, it still times out frequently.
\ours{} scales better for two reasons.
First, quantization generally benefits larger workloads more.
Second, the costs of enumerating join attribute bindings and quantization remain mostly stable:
as the size of each quantization scale is bounded by the workload size $|\queryset[\jp]|$, the cost of quantizing an attribute is only $O(\log(|\queryset[\jp]|))$ thanks to binary search.
Per-join-pattern results are provided in \reva{the full version}~\cite{fullversion}.
While the number of join patterns remains 18, the number of them where \ours{} outperforms \sequentialproc{} increases to 14 (78\%) in \texttt{job-light-1k},
17 (94\%) in \texttt{job-light-2k}, and finally 18 (100\%) in \texttt{job-light-4k}.
Overall, \ours{} demonstrates superior scalability compared with both \sequentialproc{} and \postfilter{}.

\subsection{Validation against \hybrid}
\label{subsec:exp_hybrid}

\begin{figure}[t]
    \centering
    \includegraphics[width=\columnwidth]{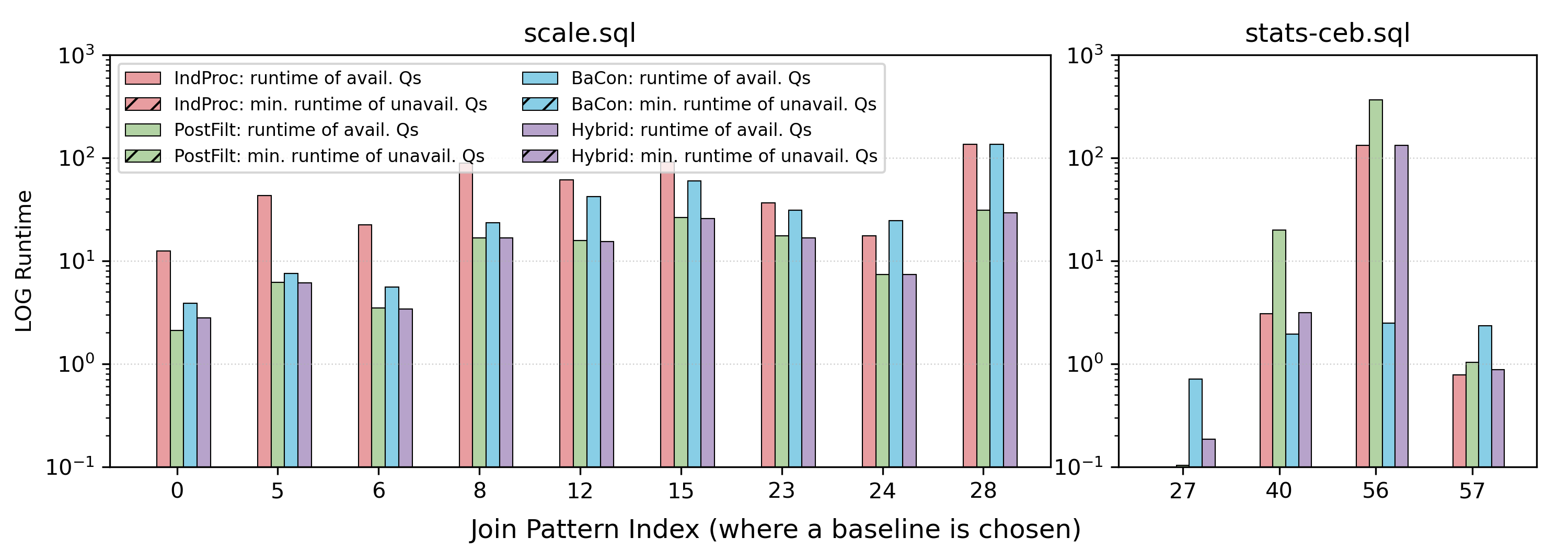}
    \vspace*{-2ex}
    \caption{\mdseries Log-scale running time per join pattern (where a baseline is chosen) in \texttt{scale} and \texttt{stats\_ceb}, including \hybrid{}'s.}
    \label{fig:hybrid_bar}
\end{figure}

As shown in \Cref{subsec:exp_per_join_pattern}, there exist join patterns for which a baseline, \sequentialproc{} or \postfilter{}, outperforms \ours{}.
These cases typically involve few queries and small intermediate join sizes.
As discussed in the full version~\cite{fullversion}, we can adopt a hybrid approach that uses a classifier to pick which method to use given a join pattern.
Here, we train \hybrid\ using observations from \texttt{job\_light\_join} and \texttt{stats\_ceb\_join}, and evaluate it on the remaining seven workloads. 
\Cref{tab:end_to_end_runtime_hybrid} reports the E2E running times of \hybrid{} (using \ours{} as the reference) on the seven test workloads.
Overall, \hybrid{} selects a baseline for 9 join patterns in \texttt{scale} and 4 join patterns in \texttt{stats\_ceb}, while selecting \ours\ for all other join patterns and workloads.
Even with \hybrid{}'s conservative design, end-to-end running times show that mispredictions can occur, sometimes catastrophically.
To take a closer look,
\Cref{fig:hybrid_bar} presents a per-join-pattern analysis, focusing on patterns where \hybrid\ chooses a baseline.
As shown in \Cref{fig:hybrid_bar}, for \texttt{scale}, all 9 switches to \postfilter{} are beneficial, yielding a total saving of 156s.
In \texttt{stats\_ceb}, \hybrid{} gains minor savings by choosing \postfilter{} on pattern 27 and \sequentialproc{} on pattern 57, but incorrectly selecting \sequentialproc{} on patterns 40 and 56 causes slowdowns of 1s and 129s, respectively.
Overall, these results confirm that \ours{} remains a safe bet for practical workloads.
With additional data statistics or training, a better \hybrid\ might be possible as future work,
but the overhead for statistics collection and training may outweigh its benefit.

\section{Related Work}\label{sec:related}

\paragraph{Select-Join-Aggregate Query Processing.} Counting join results (with selection predicates on base tables) is a special form of select-join-aggregate queries.
Standard processing typically involves pushing down selection predicates before evaluating the join and then aggregate. The classic Yannakakis algorithm \cite{10.5555/1286831.1286840} computes free-connex join-count queries in $O(N)$ time, where $N$ is the number of tuples. Recent advancements \reva{\cite{10.1145/3725241} exploit} a hybrid strategy of Yannakakis algorithm that handles acyclic join-count queries in
$\smash{O(N+N \cdot \mathsf{OUT}^{1-\frac{1}{w}})}$ time.
\ours{} draws inspiration from these techniques, specifically the recursive computation and merging of coordinate-to-count maps during backtracking from 
~\cite{10.1145/3129246,10.14778/3407790.3407797,10.14778/3342263.3342643,10.1145/3589295}. 
Cost-based optimizers in these systems, such as Free Join's~\cite{10.1145/3589295}, together with the AGM bound~\cite{atserias2017sizeboundsqueryplans,10.1145/2220357.2220363}, further inform our initial exploration of cost-based hybrid mechanism detailed in the full version~\cite{fullversion}.
However, more accurate cost estimates may conversely require more expensive estimation methods, such as \reva{LpBound}~\cite{DBLP:journals/pacmmod/ZhangMKOS25}, and/or specialized data storages and execution engines, such as Free Join's COLT~\cite{10.1145/3589295} and its integration to DuckDB~\cite{Raasveldt2022DuckDBA}.

\paragraph{Aggregation Push-down and Factorizations}
Aggregation push-down reduces intermediate results via early or deferred aggregation.
Classical approaches include eager and lazy aggregation~\cite{10.5555/645921.673154} and integrating \sql{GROUP BY} into cost-based optimization~\cite{10.5555/645920.672834}.
GuAo~\cite{10.14778/3718057.3718068} identifies cases where 
join materialization can be avoided entirely
and introduces a corresponding physical operator for SparkSQL~\cite{10.1145/2723372.2742797}.
Its propagation of frequencies grouped by join attributes resembles \ours{}'s bucketization 
in single-query settings.
However, it relies on binary joins, whereas our approach support multi-way merging of count maps.
Factorized databases~\cite{10.14778/2350229.2350242,DBLP:conf/icdt/OlteanuZ12,10.14778/2556549.2556579} eliminate tuple-level redundancy in query results through compact representations that can support efficient aggregation beyond \sql{COUNT}.
These works, focusing on eliminating intra-query redundancy, are 
complementary to our approach, which also exploits inter-query sharing.
Furthermore, these works typically employ non-traditional storage layout different from traditional relational databases.


\paragraph{Optimized Batch Processing of Aggregate Queries}
LMFAO~\cite{10.1145/3299869.3324961} evaluates batches of ground-by aggregates over shared joins without materializing intermediates by decomposing queries into views, organizing them via a join tree, and executing a multi-output plan. Like \postfilter{}, it encodes selection predicates as conditional expressions within aggregates.
Unlike \ours{}, LMFAO does not pre-process overlaps among cross-query predicates, but instead focuses on sharing computation 
of join tree traversal and attribute-ordered evaluation,
and is implemented as a standalone execution engine.



\paragraph{Multi-Query \& Shared Workload Optimization}
Multi-query optimization (MQO)~\cite{10.1145/42201.42203} identifies shared/similar subexpressions across queries and 
evaluate them jointly or via plan rewriting.
Zhou et al.~\cite{10.1145/1247480.1247540} propose a 
common subexpression manager integrated into Microsoft SQL Server,
demonstrating improvements on workloads of tens of queries. 
Shared Workload Optimization (SWO)~\cite{10.14778/2732279.2732280} extends this idea to larger workloads by sharing operators, often requiring specialized execution engines.
These approaches target general workloads and require accurate cost estimates.


\paragraph{\revc{Learned Cardinality Estimation}}
\revc{
    Query-aware cardinality estimation (CE) models~\cite{park2020quicksel,10.14778/3329772.3329780,10.14778/3476249.3476254,wu2021unified,kipf2018learned,10.14778/3476249.3476259,DBLP:conf/edbt/MullerWL23,10.1145/3514221.3517896,10.1145/3514221.3526179,10.14778/3626292.3626302} learn 
    from query-count pairs.
    Training is typically performed either in a single-shot manner, where a model is (re-)trained from scratch on a batch of queries~\cite{10.14778/3329772.3329780, 10.1145/3514221.3526179,li2021cardinality,10.14778/3476249.3476254,10.1145/3514221.3517896}, or in multiple rounds, where the model is incrementally updated
    as new query-count pairs become available~\cite{10.14778/3626292.3626302,10.1145/3654932,park2020quicksel}.
    Some hybrid CE models~\cite{kipf2018learned,10.14778/3626292.3626302} additionally require data statistics, e.g., samples/histograms.
    Recent work studies robustness, training efficiency, and adaptivity to data updates and workload shifts~\cite{10.1145/3588713,10.14778/3725688.3725708,jgmp,10.1145/3514221.3526179,10.1145/3639293,10.14778/3583140.3583164,10.1145/3654932,10.14778/3626292.3626302}, increasing the demand for efficiently supporting counting queries.
}

\paragraph{\common{Other Related Work}}
\common{
    Our work is also related to continuous query processing.
    More details are provided in the full version~\cite{fullversion}.
}
\section{Conclusion}
\label{sec:conclusion}

In this paper, we presented \ours, a method for efficient batch processing of counting queries.
\ours\ brings together multiple optimization ideas,
focusing particularly on developing compact, alternative representations of intermediate results that enable fine-grained sharing of computation.
\ours\ combines lightweight SQL execution with client-side processing, 
incorporating a suite of optimizations to ensure practical performance on our target workloads.
A strength of \ours\ is its practicality: \reva{it can be deployed without modifying DBMS internals or physical designs.
Components of \ours\ can be accelerated using efficient UDF implementations (such as C-extensions in PostgreSQL), if they are supported by the DBMS.}
Our results demonstrate significant performance gains across real workloads with diverse characteristics.
This level of improvement is empowering:
shorter (re)training times make learned CE more practical and allow larger, more comprehensive training workloads.

\ours\ opens up several promising directions for future work.
Extending it to support richer join structures, more expressive selection predicates,
and more complex aggregates (complex expression inputs or non-algebraic aggregate functions) would further broaden its applicability.
Beyond functional extensions, there are opportunities for more optimization, such as
sharing computation across join patterns;
delayed evaluation of $\oplus$ and $\otimes$ in count map expressions (and adapting \ComputeCounts\ to take advantage);
more intelligent selection of tree-based execution plans;
aggressive ``short-cutting'' of all FK-PK edges;
caching and more sophisticated parallelization.
Finally, new applications and additional optimizations will likely require models beyond the current \hybrid\ to enable cost-based selection of processing methods.

\begin{acks}
    This work of\ J.Y., P.A., and Y.L. was partially supported by NSF Grant IIS-2402823. 
    P.A. was also supported by NSF Grant CCF-2223870 and a US-Israel Binational Science Foundation Grant 2022131. 
    This work of X.H. was supported by the Natural Sciences and Engineering Research Council of Canada Discovery Grant.
\end{acks}

\clearpage

\balance
\bibliographystyle{ACM-Reference-Format}
\bibliography{sample}

\clearpage

\begin{appendix}
    \section{Notation Table}

\begin{table}[h]
    \caption{\mdseries \revc{Table of Notations.}}
    \label{tab:notations}
    \vspace{-1em}
    \small
    \begin{tabular}{l|p{0.65\columnwidth}}
        Notation & Description
        \\\hline\hline
            $\database$ & Database
        \\
            $\tabs$ & A (sub)set of tables in $\database$
        \\
            $\tab$ & A table in $\database$
        \\
            \hspace*{1em}$\Attrs{\tab}$ & Set of attributes of table $\tab$   
        \\\hline
            $\queryset$ & Set of all counting queries to evaluate over $\database$
        \\
            $\query$ & A counting query
        \\
            \hspace*{1em}$\Tables{\query}$ & Set of tables referenced by $\query$
        \\
            \hspace*{1em}$\JPreds{\query}$ & $\query$'s join predicates
        \\
            \hspace*{1em}$\SPreds{\query}$ & $\query$'s selection predicates
        \\
            \hspace*{2em}$\SPreds{\query}(A)$ & query range for selection attribute $A$
        \\
            \hspace*{1em}$\JAttrs{\query}$ & Set of attributes referenced by $\query$'s join predicates 
        \\
            \hspace*{2em}$\JAttrs{\query}[\tabs|\tabs']$ & Subset of attributes of $\tabs$ used by $\query$ to join with $\tabs'$,
                where $\tabs$ and $\tabs'$ are a pair of disjoint subsets of tables in $\Tables{\query}$
        \\
            \hspace*{1em}$\SAttrs{\query}$ & Set of attributes referenced by $\query$'s selection predicates
        \\
            \hspace*{2em}$\SAttrs{\query}[\tabs]$ & Subset of attributes of $\tabs$ referenced by $\query$'s selection predicates,
                where $\tabs$ is a subset of tables in $\Tables{\query}$
        \\\hline
            $\jp = \langle \Tables{\jp}, \JPreds{\jp} \rangle$ & A join pattern characterized by a set tables $\Tables{\jp}$
                joined by predicates $\JPreds{\jp}$ (with no selection predicates)
        \\
            \hspace*{1em}$\queryset[\jp]$ & Subset of queries in $\queryset$ with join pattern $\jp$
        \\
            \hspace*{1em}$\SAttrs{\jp}$ & Set of attributes involved in selection predicates in all queries of $\queryset[\jp]$
        \\
            \hspace*{1em}$\SPreds{\jp}(A)$ & Set of query ranges associated with selection attribute $A$ in all queries of $\queryset[\jp]$
        \\\hline\hline
            $\QuantScales_\jp$ & The collection of all quantization scales for queries in $\queryset_\jp$,
                one for each selection attribute in $\SAttrs{\jp}$,
                inducing a grid over a ...
        \\
            \hspace*{1em}$\QuantScales_\jp[\tab]$ & Subset of quantization scales for selection attributes in $\SAttrs{\jp}$ that belong to table $R$
        \\
            \hspace*{1em}$\QuantScale_\attr \in \QuantScales_\jp$ & Quantization scale for attribute $\attr \in \SAttrs{\jp}$
        \\
            \hspace*{2em}$\QuantScale_\attr(x)$ & Serial integer id of the bucket containing $x$
                (or $0$ if $x$ lies outside all buckets), where $x$ is a value from $\attr$'s domain
        \\
            \hspace*{1em}$\myvec{b}$ & A grid coordinate, with one integer component per selection attribute
        \\\hline
            $\CountMap$ & A count map
        \\
            \hspace*{1em}$\CountMap[\myvec{b}]$ & The count of tuples falling within the grid cell with coordinate $\myvec{b}$
        \\\hline\hline
            $\beta$ & Parameter used by optimized \ourrecursive\ to batch \ProcessTable\ calls (\Cref{sec:implementation:sql})
        \\\hline\hline 
    \end{tabular}
\end{table}

    \section{More Details of Basic \ours{}}
We present the detailed construction algorithm of $\QuantScale_{\attr}$ for a selection attribute $\attr{}$ in \Cref{subsec:appendix_subsec_buckets}.

\subsection{Construction of Buckets}\label{subsec:appendix_subsec_buckets}
Continuing from \Cref{sec:ours:quantization}, we extract from $\queryset[\jp]$ the set $\Delta = \{ \SPreds{\query}(\attr) \mid \exists \query \in \queryset[\jp]: \attr \in \SPreds{\query} \}$ of predicate ranges associated with $\attr$.
We transform each range $G$ specified by $\SPreds{\query}(\attr)$ to a left-closed, right-open interval.
Concretely, a left-open boundary ($l$) is converted to a left-closed boundary $\narrow{succ}(l)$, and a right-closed boundary ($r$) is converted to a right-open boundary $\narrow{succ}(r)$, where $\narrow{succ}(\cdot)$ advances the value by one unit at the attribute's resolution
\footnote{For example, if $\alpha$ is an integer, $\nxt{(\alpha)} = \alpha + 1$; if $\alpha$ is a \sql{timestamp without time zone}, $\nxt{(\alpha)} = \alpha + 1$ microsecond; and so on.}
.
Next, we sort the boundaries in ascending order and remove duplicates. 
The $i$-th boundary and the $(i + 1)$-th one (if exists), where there are more left boundaries than right ones among the first $i$ boundaries, form a left-closed-right-open interval in the domain of $\attr$ that is of interest to at least one query. 
We refer to each of these intervals as a bucket, and all these buckets form $\QuantScale_{\attr}$.
As detailed in \Cref{sec:ours:quantization},
a value $v$ is quantized into an integer, i.e., the id of the bucket containing $v$, or 0 if $v$ lies outside all of the buckets in $\QuantScale_{\attr}$.
Note that if a query in $\jp$ has no selection predicates on $\attr{}$, it still satisfies \Cref{lemma:quantization} by setting $i_1 = 0$ and $i_2 = |\QuantScale_{\attr}|$.

We present an example of quantization in \Cref{example:quantization}.

\begin{example}\label{example:quantization}
    Continuing from \Cref{sql:stats_ceb_example}, we have
    
    $$\QuantScale_{\sql{c.Score}} = \{[0, 1)\}$$,
    
    $$\QuantScale_{\sql{ph.PostHistoryTypeId}} = \{[1, 2)\}$$, and
    
    $$\QuantScale_{\sql{ph.CreationDate}} = \{[\sql{2010-09-14 11:59:07}, \infty)\}$$.
    
    By \Cref{lemma:quantization}, the integer range associated with $\query_3$ is $[1, 1]$ for \sql{c.Score}, $[1, 1]$ for \sql{ph.PostHistoryTypeId}, and $[0, 1]$ for \sql{ph.Creat} \sql{-ionDate}.
    Similarly, the integer range associated with $\query_4$ is $[0, 1]$ for \sql{c.Score}, $[1, 1]$ for \sql{ph.PostHistoryTypeId}, and $[1, 1]$ for \sql{ph.Creat} \sql{-ionDate}.
\end{example}
    \clearpage
    \section{More Details of \ours{} Implementation}
We present the details of \OptProcessTable{} and \optourrecursive{} in \Cref{subsec:appedix_subsec_opt}.

\subsection{Details of \OptProcessTable{} and \optourrecursive{}}\label{subsec:appedix_subsec_opt}

\begin{sqllisting}[float=t]{\mdseries SQL code for $\OptProcessTable(\tab, U, \attrs^\narrow{in \& out})$.}{sql:opt-count-table}
`\textnormal{\textbf{Input:} Set $U$ of mappings, with each binds attributes $\Jin_1, \Jin_2, \ldots$ to specific values, and $\attrs^\narrow{in \& out} = \{\Jin_1, \Jin_2, ..., \Jout_1, \Jout_2, \ldots\}$ specifies the set of attributes for partitioning result entries. The quantization scales are $\QuantScales_\jp[\tab] = \{ \QuantScale_{S_1}, \QuantScale_{S_2}, \ldots \}$.}`
`\textnormal{\textbf{Output:} Result entries have the form $\langle v, \myvec{b}, c\rangle$, sorted by $v$, where $v$ is a mapping from $\attrs^\narrow{in \& out}$ to values, $\myvec{b}$ is a grid coordinate for $\QuantScales_\jp[\tab]$, and $c$ is the associated count.}`
`\noindent\rule{0.98\columnwidth}{0.2pt}`
SELECT $\Jin_1$, $\Jin_2$, ..., $\Jout_1$, $\Jout_2$, ..., B1, B2, ..., COUNT(*)
FROM (
    SELECT $\Jin_1$, $\Jin_2$, ..., $\Jout_1$, $\Jout_2$, ...,
        quantize($S_1, \QuantScale_{S_1}$) AS B1, quantize($S_2, \QuantScale_{S_2}$) AS B2, ...
    FROM $\tab$
    WHERE $\Jin_1$ BETWEEN $\min(U(\Jin_1))$ AND $\max(U(\Jin_1))$
          $\Jin_2$ BETWEEN $\min(U(\Jin_2))$ AND $\max(U(\Jin_2))$ AND ...
) AS TMP
GROUP BY $\Jin_1$, $\Jin_2$, ..., $\Jout_1$, $\Jout_2$, ..., B1, B2, ...
ORDER BY $\Jin_1$, $\Jin_2$, ..., $\Jout_1$, $\Jout_2$, ...;
\end{sqllisting}

\newcommand{\dict}{\ensuremath{\mathbf{M}}}
\newcommand{\vin}{\ensuremath{v^{\narrow{in}}}}
\newcommand{\vout}{\ensuremath{v^{\narrow{out}}}}
\newcommand{\win}{\ensuremath{\omega^{\narrow{in}}}}
\newcommand{\wout}{\ensuremath{\omega^{\narrow{out}}}}
\begin{algorithm}[b]
\caption{$\optourrecursive(\tab, U)$}\label{alg:our_opt_recursive}
\begin{algorithmic}[1]
\Require{Set $U$ of mappings, with each binds a subset of $\tab$'s attributes to specific values.
    Implicitly, the function also has access to $\database$, $\jp$ and its plan tree, and the quantization scales $\QuantScales_\jp$.}
\Ensure{Dictionary $\dict$ maps each binding $u \in U$ to a count map $\CountMap$ over $\subtree(\tab)$; formally, $\dict \coloneq \{u \mapsto \CountMap~\mid~u \in U\}$.
    Each $\CountMap$ (or equivalently $\dict[u]$) is complete w.r.t.\ result tuples of $\queryset[\jp]$ restricted to $\subtree(\tab)$ and consistent with $u$.}
\State $\dict \gets \{u \mapsto~\text{an empty count map}~~\mid~\forall u \in U\}$; 
\State $n_{\text{subsequence}} \gets 0$; \Comment{counter for the number of subsequences to be processed}
\For{each $\mathbf{S}[v]$ of entries of the form $\langle v, \cdot, \cdot \rangle$ with the same $v$,
    returned by $\OptProcessTable\big(\tab, U, \JAttrs{\jp}(\parent(\tab)\,|\,\tab) \cup \JAttrs{\jp}(\tab\,|\,\children(\tab))\big)$
}\label{alg:our_opt_recursive:loop}
    \State $\vin \gets \langle v[\Jin_1], \myvec[\Jin_2], \dots \rangle$;
    $\vout \gets \langle v[\Jout_1], \myvec[\Jout_2], \dots \rangle$;\label{alg:our_opt_recursive:postcheck}
    \If{$\vin \notin U$} \Comment{the false positive case mentioned in \Cref{sec:implementation:sql}: join attribute values lie in $[\min(U), \max(U)]$ but are not contained in $U$}
        \State \textbf{continue};
    \EndIf\label{alg:our_opt_recursive:postcheck-end}
    \State $n_{\text{subsequence}} \gets n_{\text{subsequence}} + 1$;
    \If{$n_{\text{subsequence}}$=$\batchparam~$ or $~\mathbf{S}[v]$ is the last subsequence returned by $\OptProcessTable$} \Comment{batch processing of $\batchparam$ subsequences; equivalently, batch processing a set $V$ of consecutive values of $v$}\label{alg:our_opt_recursive:batch}
        \For{each table $\tab_i \in \children(\tab)$}\label{alg:our_opt_recursive:recursive_loop}
            \State $U_i \gets \emptyset$;
            \For{each subsequence $\mathbf{S}[\omega]$ in the batch} \Comment{each subsequence in the batch ended with $\mathcal{S}[v]$}\label{alg:our_opt_recursive:binding-collection}
                \State $\wout \gets \langle \omega[\Jout_1], \omega[\Jout_2], \dots \rangle$; 
                \State $u_i \gets \left\{ A' \mapsto \wout(A) \;\middle|\;
                    {\footnotesize
                    \begin{aligned}
                        & A \in \JAttrs{\jp}(\tab | \tab_i)\\ 
                        & \land A' \in \JAttrs{\jp}(\tab_i | \tab)\\
                        & \land \JPreds{\jp} \Rightarrow (A = A')
                    \end{aligned}}
                    \right\}$;\label{alg:our_opt_recursive:project_binding}
                \State $U_i \gets U_i \cup u_i$;
            \EndFor\label{alg:our_opt_recursive:binding-collection-end}
            \State $\dict_i \gets \optourrecursive(\tab_i, U_i)$;\label{alg:our_opt_recursive:child-dict}
        \EndFor
        \For{each subsequence $\mathbf{S}[\omega]$ in the batch}\label{alg:our_opt_recursive:contribute-to-dict}
            \State $\win \gets \langle \omega[\Jin_1], \omega[\Jin_2], \dots \rangle$; $\wout \gets \langle \omega[\Jout_1], \omega[\Jout_2], \dots \rangle$; 
            \State $\CountMap_0 \gets \{ \myvec{b} \mapsto c \mid \langle \omega, \myvec{b}, c \rangle \in \mathbf{S}[\omega]\}$;
            \For{each table $\tab_i \in \children(\tab)$}
                \State $\CountMap_0 \gets \CountMap_0 \otimes \dict_i[u_i]$;\Comment{$u_i$ as defined in Line~\ref{alg:our_opt_recursive:project_binding}}\label{alg:our_opt_recursive:multiply}
            \EndFor
            \State $\dict[\win] \gets \dict[\win] \oplus \CountMap_0$;\label{alg:our_opt_recursive:add}
        \EndFor\label{alg:our_opt_recursive:contribute-to-dict-end}
        \State $n_{\text{subsequence}} \gets 0$;
    \EndIf\label{alg:our_opt_recursive:batch-end}
\EndFor\label{alg:our_opt_recursive:loop-end}
\State \Return $\dict$;
\end{algorithmic}
\end{algorithm}

In \OptProcessTable{} (\Cref{sql:opt-count-table}), the second and third arguments differ from those of \ProcessTable.
Specifically, \OptProcessTable{} takes a set $U$ of bindings rather than a single binding $u$, and it is passed both $\Jin_1, \Jin_2, \dots$ and $\Jout_1, \Jout_2, \dots$ to enable partition of result entries.
Moreover, in the inner block, $\Jin_1, \Jin_2, \dots$ are filtered using safe but potentially imprecise range bounds.
Concretely, $\min{U(\Jin_i)}$ (resp. $\max{U(\Jin_i)}$) denotes the minimum (resp. maximum) value in the set $\{u(\Jin_i)~\mid~\forall u \in U\}$.
Finally, \OptProcessTable{} also \sql{SELECT}s $\Jin_1, \Jin_2, \dots$ and partitions (and orders) the result entries by these attributes, allowing the receiver, \optourrecursive{}, to discard tuples that do not correspond to any binding in $U$.

We propose \optourrecursive{} (\Cref{alg:our_opt_recursive}) by incorporating batching into \ourrecursive{}.
Instead of producing a single count map, \optourrecursive{} outputs a dictionary that maps each binding $u \in U$ to a corresponding count map, enabling batch processing of \batchparam{} consecutive bindings.
Lines \ref{alg:our_opt_recursive:loop}-\ref{alg:our_opt_recursive:loop-end} enumerate subsequences of result entries.
Unlike \ourrecursive{}, these subsequences are not processed immediately; computation is deferred until either $\batchparam{}$ subsequences have been accumulated or the end of the returned entries is reached (Line \ref{alg:our_opt_recursive:batch}).
As discussed earlier, returned tuples may not belong to $U$.
Therefore, a post-check filters tuples by testing whether $v$'s projection onto $\Jin_1, \Jin_2, \dots$ is contained in $U$ (Lines \ref{alg:our_opt_recursive:postcheck}-\ref{alg:our_opt_recursive:postcheck-end}).
During batch processing (Lines \ref{alg:our_opt_recursive:batch}-\ref{alg:our_opt_recursive:batch-end}), we first collect the set $U_i$ of distinct bindings on the join attributes between the child table $\tab_i$ and $\tab{}$ (Lines \ref{alg:our_opt_recursive:binding-collection}-\ref{alg:our_opt_recursive:binding-collection-end}).
We then recursively invoke \optourrecursive{} for each pair ($\tab_i, U_i$). 
Finally, Lines \ref{alg:our_opt_recursive:contribute-to-dict}-\ref{alg:our_opt_recursive:contribute-to-dict-end} update the dictionary $\dict{}$ by aggregating the contribution of each subsequence.
The procedure for merging count maps largely follows \ourrecursive{}, with two key differences: $\tab_i$'s count map is indexed by $u_i$ and obtained from the returned dictionary; and the merged count maps are accumulated into the dictionary entry indexed by $\win{}$.
Note that $\win{}$ (and similarly $\vin{}$ in Line \ref{alg:our_opt_recursive:postcheck}) is empty for the root table, which has no $\Jin$ attributes.

\subsection{Choosing among \ours/\sequentialproc/\postfilter}
\label{sec:implementation:hybrid}

We introduce an approach called \emph{\hybrid} for picking the appropriate method to use among the three for a given query pattern.
As we will see in \Cref{subsec:exp_hybrid}, \ours\ performs well across our target workloads,
so \hybrid\ only serves to validate the robustness of \ours.
Nonetheless, we briefly describe \hybrid\ here for completeness.

\begin{figure}[t]
    \centering
    \includegraphics[width=0.95\columnwidth]{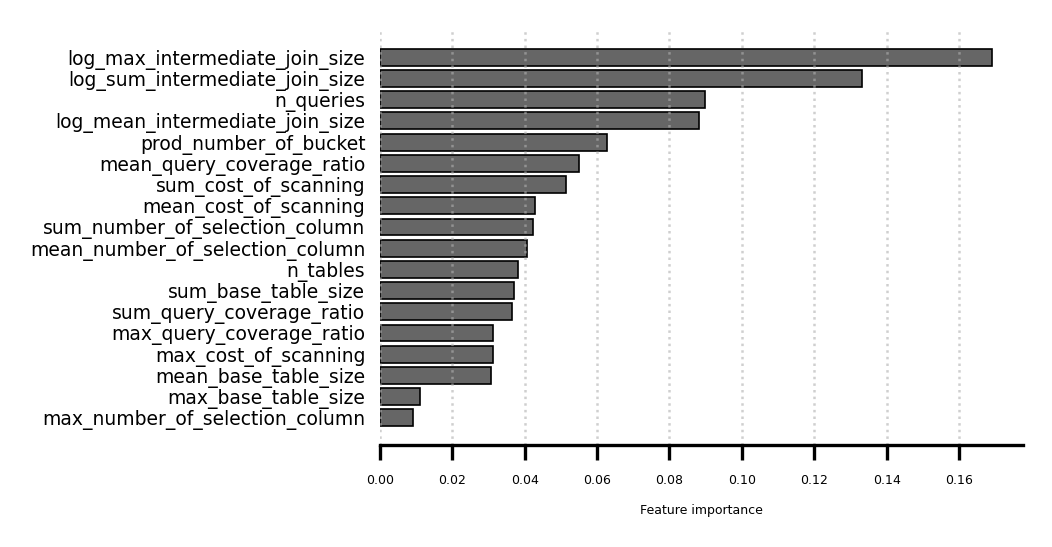}
    \vspace*{-5ex}
    \caption{\mdseries Features used by \hybrid, and their importance in a model trained as described in \Cref{subsec:exp_hybrid}.}
    \label{fig:feature_importance}
    \vspace*{-4ex}
\end{figure}

We train \hybrid\ from sample workloads as a classifier that predicts, given a join pattern,
one of three classes $\{0, 1, 2\}$ representing \ours, \sequentialproc, and \postfilter\ respectively.
We use a lightweight model based on \texttt{RandomForestClassifier}~\cite{randomforestclassifier} over a suite of features --- the most important of which are illustrated in \Cref{fig:feature_importance}.
Features with prefixes \sql{sum\_}, \sql{max\_}, \sql{mean\_}, and \sql{prod\_} aggregate a collection of values:
for example, \sql{prod\_number\_of\_bucket} is the product of the number of buckets across all quantization scales for a join pattern;
\sql{mean\_query\_coverage\_ratio} is the average, taken per query, of the ratio between the number of grid cells queried and \sql{prod\_number\_of\_bucket}.
Features containing \sql{intermediate\_join\_size} are AGM bounds~\cite{atserias2017sizeboundsqueryplans,10.1145/2220357.2220363}
(i.e., overestimation) of intermediate result sizes of join subplans chosen by PostgreSQL for joining all tables in the join pattern.
Features with \sql{cost\_of\_scanning} denote PostgresSQL's estimated cost for enumerating a table ordered by join attributes that connects it to its descendant tables (if any), reflecting a key component of \ours{}.
In general, we include as features per-join-pattern statistics that are inexpensive to collect at runtime.
However, we exclude more advanced selectivity estimates provided by PostgreSQL,
because we have found them to be unreliable and systematically underestimating the costs of baselines.

To ensure a lightweight model and fast inference, we train our model with \sql{train\_val\_split}=0.25, \sql{n\_estimators}=100, \sql{min\_samples\_le} \sql{-af}=2, and \sql{class\_weight}=\{``0'': 1, ``1'': 2, ``2'': 2\}.
The higher weights for classes 1 and 2 penalize mistakenly picking baseline approaches,
because such mistakes cost much higher.
To obtain class labels for training data, we measure and compare the running time for all three methods.
We deem a baseline method \emph{safe} if its running time is either (i) at least $1.5\times$ faster than \ours, or (ii) no less than $10$ seconds faster than \ours.
We assign class $1$ only when \sequentialproc\ is safe and \postfilter\ is either unsafe or slower;
class $2$ is assigned analogously;
all remaining cases are labeled as class $0$.
For inference, rather than selecting the class with the highest posterior probability,
we require the predicted probability of choosing \sequentialproc\ or \postfilter\ to exceed a precision threshold ($0.9$ by default),
calibrated to guarantee minimum precision on held-out validation data.
If no probability threshold can guarantee the required precision, the one with highest precision is selected.
This robust thresholding strategy deliberately trades recall for safety, selecting baselines only under high confidence.
    \clearpage
    \section{More Details of Experiments}
We present the results of materialized-view-based postfiltering in \Cref{subsec:appendix_subsec_exp_materialized_view}, 
the results of multi-threaded execution in \Cref{subsec:appendix_subsec_exp_multi_thread}, 
and the results of varying \batchparam{} in \Cref{subsec:appendix_subsec_exp_batch_param}.
We present the per-join-pattern results of both the remaining real workloads and scalability workloads in \Cref{subsec:appendix_subsec_exp_per_jp}.
Additionally, the most detailed result of \ours{} and \postfilter{} can be found in the log files of the form \texttt{results/*.print} in our repository~\cite{bacon_repo}.

\revb{
\subsection{Experiments of Materialized-View-Based Postfiltering}\label{subsec:appendix_subsec_exp_materialized_view}
An alternative implementation of \postfilter{} materializes the unfiltered join for each join pattern and performs aggregations over it.
However, this approach is impractical for join patterns with very large unfiltered join sizes, plus this information is not free, due to prohibitive storage overhead.
Therefore, we evaluate the materialized-view-based variant only on DSB's \texttt{dsb-grasp-20k} and JOB's \texttt{synthetic}, where the materialized view cardinalities remain manageable and \postfilter{} outperforms \sequentialproc{} (as shown in \Cref{tab:end_to_end_runtime}).
The result is presented in \Cref{tab:end_to_end_materialized}.
}

\begin{table}[h]
    \caption{\revb{\mdseries End-to-end running times (in seconds, rounded) of the materialized-view-based algorithm, compared to \postfilter{}.}}
    \label{tab:end_to_end_materialized}
    \vspace{-1em}
    \small
    \begin{tabular}{r||r|r|r|}
            Query Workload & \begin{tabular}{@{}r@{}}View \\ Cardinalities\end{tabular} & \postfilter{} & \begin{tabular}{@{}r@{}}Materialized-view\\-based Postfiltering\end{tabular}
        \\\hline\hline
            dsb-grasp-20k & $< 6 \cdot 10^6$ & 4,211 & 4,356
        \\\hline\hline
            synthetic & $< 5 \cdot 10^8$ & 11,329+ & 12,264+
        \\\hline
    \end{tabular}
\end{table}

\revc{
\subsection{Experiments of Multi-threaded Execution}\label{subsec:appendix_subsec_exp_multi_thread}
We reports results for \sequentialproc{}, \postfilter{}, and \ours{} under multi-threaded execution.
For the baselines, \sequentialproc{} and \postfilter{}, we enable PostgreSQL's parallel execution by setting \sql{max\_parallel\_workers} to 8 and \sql{max\_parallel\_workers\_per\_gather} to 4. 
Since all their computation is performed within the database server, they can fully exploit this parallelism.
In contrast, without an elaborate parallelization design, \ours{} uses 4 processes, each handling a subset of join patterns assigned in a round-robin manner, while keeping the PostgreSQL server in single-threaded mode.
The results are presented in \Cref{tab:end_to_end_multi_threaded}.
While we adopt this simple, join-pattern-level parallelization, more sophisticated strategies for \ours{} are possible (e.g., parallel merging of count maps from descendants or cost-based partitioning of join patterns), while we leave as future work.
}

\begin{table}[h]
    \caption{\revc{\mdseries Multi-threaded mode: end-to-end running times (in seconds, rounded) and relative speedups. +: lower-bound values, as \sequentialproc{} is capped at 900s per query and \postfilter{} at 3{,}600s per join pattern.}}
    \label{tab:end_to_end_multi_threaded}
    \vspace{-1em}
    \small
    \begin{tabular}{c||r|r|r|r|r|}
          Query Workload & \begin{tabular}{@{}r@{}}\sequentialproc{}\\(parallel)\end{tabular} & \begin{tabular}{@{}r@{}}\postfilter{}\\(parallel)\end{tabular} & \begin{tabular}{@{}r@{}}\ours{}\\(parallel)\end{tabular} & \begin{tabular}{@{}r@{}}Speedup \\vs. \sequentialproc{}\end{tabular}
           \\\hline\hline
            synthetic & 5,271 & 2,565 & 531 & $\uparrow$ 9.92 $\times$
            \\\hline
            scale & 11,003+ & 14,715+ & 1,097 &  $\uparrow$ 10.03+ $\times$
            \\\hline
            job-light & 540 & 8,046 & 273 & $\uparrow$ 1.98+ $\times$
            \\\hline
            job-light-single & 66 & 3 & 11 & $\uparrow$ 5.83 $\times$
            \\\hline
            job-light-join & 2,595 & 11,198+ & 510 &$\uparrow$ 5.09 $\times$
            \\\hline
            stats-ceb & 3,390+ & 43,661+ & 19 & $\uparrow$ 180.79+ $\times$
            \\\hline
            stats-ceb-single & 17 & 2 & 2 & $\uparrow$ 8.48 $\times$
            \\\hline
            stats-ceb-join & 32,384+ & 51,120+ & 54 & $\uparrow$ 598.48+ $\times$
            \\\hline
            dsb-grasp-20k & 6,241 & 913 & 359 & $\uparrow$ 17.37 $\times$
         \\\hline
    \end{tabular}
\end{table}

\subsection{Experiments of Various \batchparam{}}\label{subsec:appendix_subsec_exp_batch_param}
We report results for \batchparam{} $=$ 5,000, 50,000 and 500,000 respectively, with \batchparam{}$=$ 50,000 as the default setting used throughout \Cref{sec:experiment}.
The corresponding end-to-end running times are shown in \Cref{tab:end_to_end_batchparam}.
Overall, increasing \batchparam{} reduces the end-to-end running time, at the cost of higher space overhead (e.g., larger $\dict_i$ in Line \ref{alg:our_opt_recursive:child-dict} of \Cref{alg:our_opt_recursive}).
As indicated by the last two columns of \Cref{tab:end_to_end_batchparam}, the primary source of performance differences lie in the number of SQLs issued to the underlying database and the time spent fetching their results.
With larger \batchparam{}s, \ours{} issues fewer SQLs and incurs lower result-fetching overhead, resulting in more savings on the end-to-end running times. 

\begin{table}[h]
    \caption{\mdseries \texttt{job-light}: end-to-end running times (in seconds, rounded), together with the total number of SQLs executed on the underlying database and the total time spent fetching their results, for different values of \batchparam{}. The default configuration in \Cref{sec:experiment} (\batchparam{} = 50,000) is shown in bold.}
    \label{tab:end_to_end_batchparam}
    \vspace{-1em}
    \small
    \begin{tabular}{r||r|r|r|}
            \batchparam{} & E2E running time & Total \# of SQLs & \begin{tabular}{@{}r@{}}Total result-fetching\\ overhead\end{tabular}
        \\\hline\hline
            5,000 & 922 & 19,427 & 639
        \\\hline\hline
            \textbf{50,000} & \textbf{728} & \textbf{1,974} & \textbf{497}
        \\\hline\hline
            500,000 & 584 & 243 & 288
         \\\hline
    \end{tabular}
\end{table}

\subsection{More Per-Join-Pattern Results}\label{subsec:appendix_subsec_exp_per_jp}
\begin{figure*}[htbp]
    \centering
    \includegraphics[width=\columnwidth]{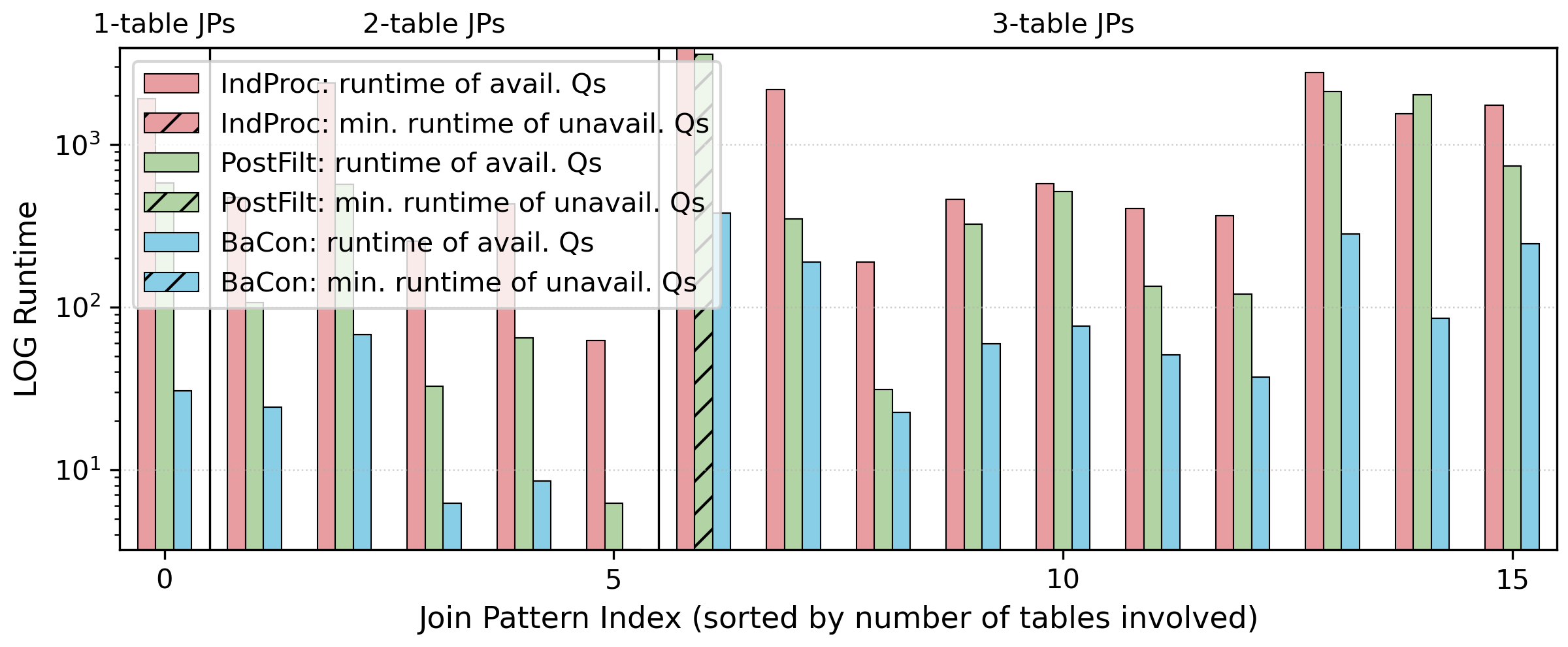}
    \vspace*{-2ex}
    \caption{\mdseries Log-scale running time per join pattern in \texttt{synthetic}.}
    \label{fig:synthetic_bar}
\end{figure*}

\begin{figure*}[htbp]
    \centering
    \includegraphics[width=0.25\columnwidth]{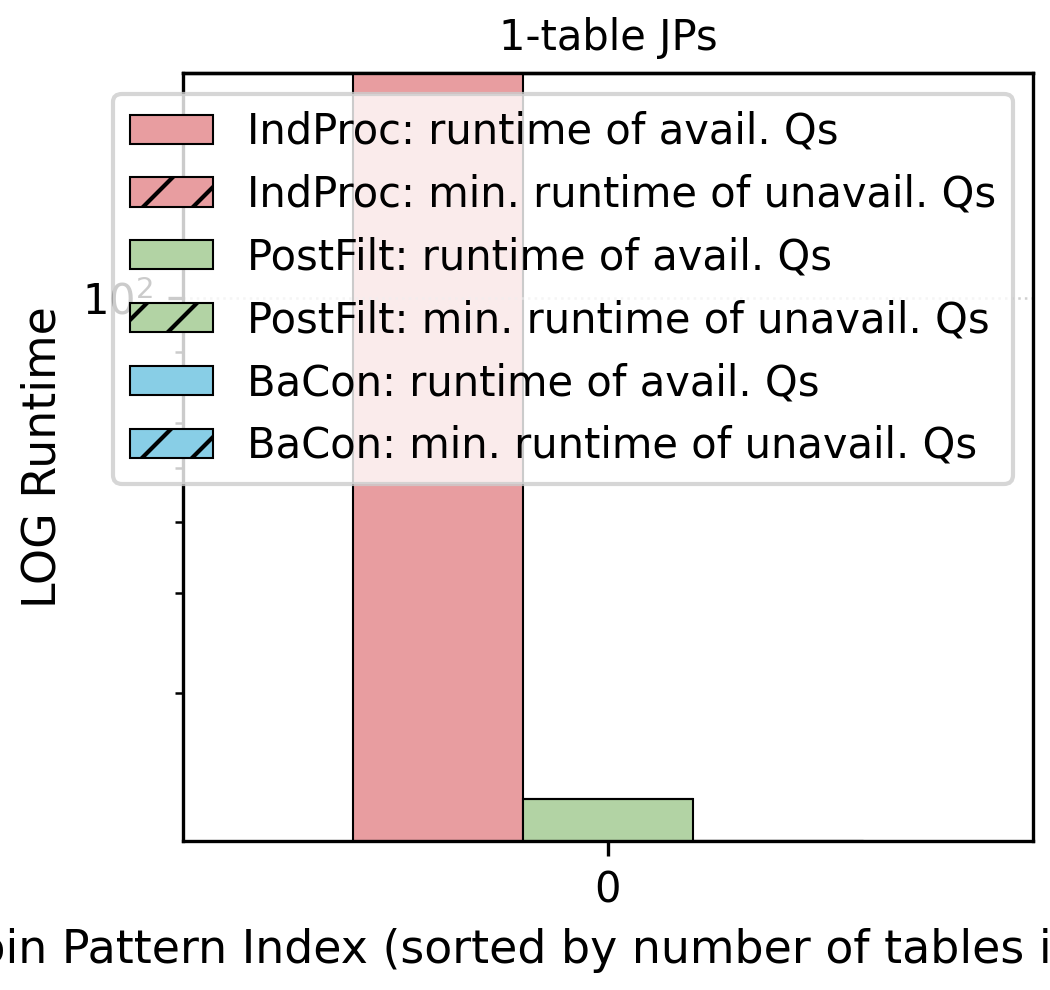}
    \vspace*{-2ex}
    \caption{\mdseries Log-scale running time per join pattern in \texttt{job\_light\_single}.}
    \label{fig:job_light_single_bar}
\end{figure*}

\begin{figure*}[htbp]
    \centering
    \includegraphics[width=\columnwidth]{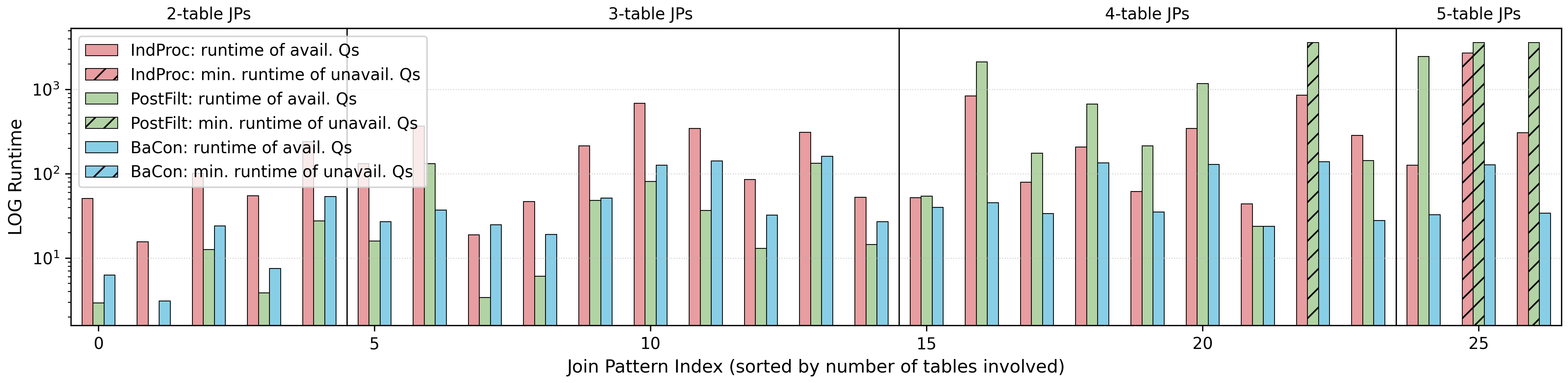}
    \vspace*{-2ex}
    \caption{\mdseries Log-scale running time per join pattern in \texttt{job\_light\_join}.}
    \label{fig:job_light_join_bar}
\end{figure*}

\begin{figure*}[htbp]
    \centering
    \includegraphics[width=\textwidth]{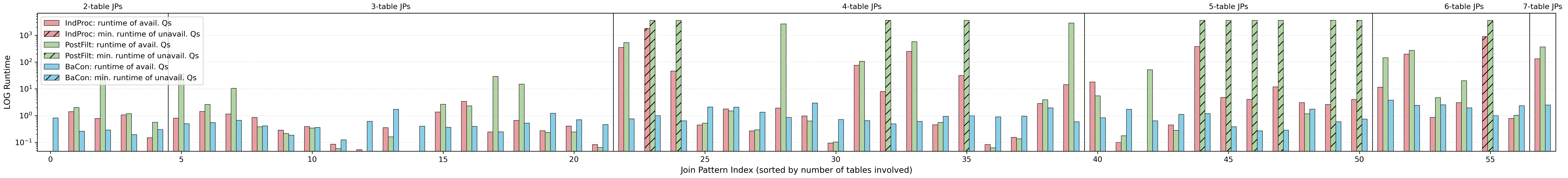}
    \vspace*{-2ex}
    \caption{\mdseries Log-scale running time per join pattern in \texttt{stats\_ceb}.}
    \label{fig:stats_ceb_bar}
\end{figure*}

\begin{figure*}[htbp]
    \centering
    \includegraphics[width=0.25\columnwidth]{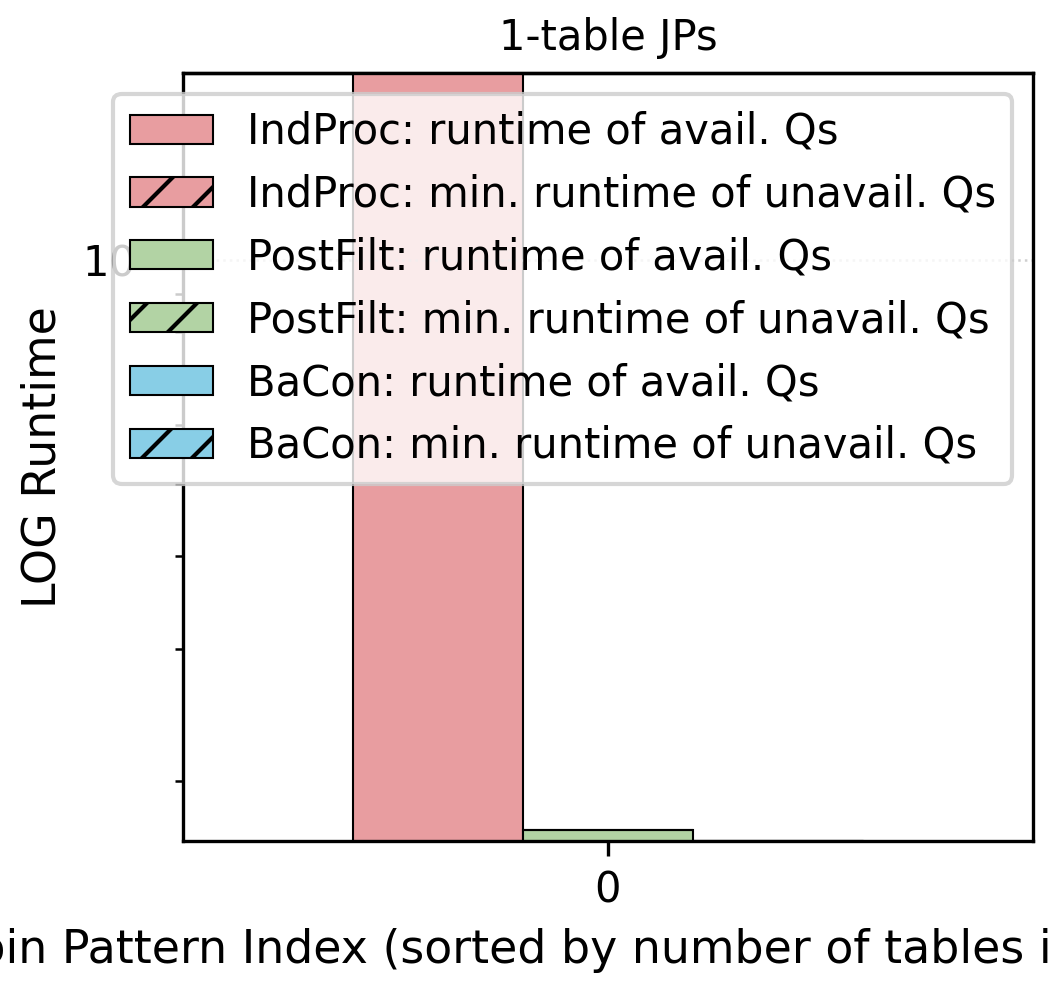}
    \caption{\mdseries Log-scale running time per join pattern in \texttt{stats\_ceb\_single}.}
    \label{fig:stats_ceb_single_bar}
\end{figure*}

\begin{figure*}[htbp]
    \centering
    \includegraphics[width=\textwidth]{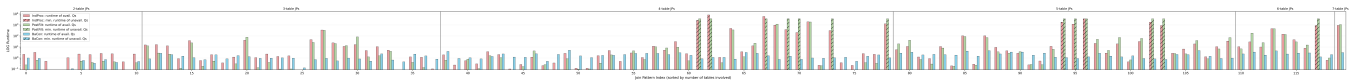}
    \vspace*{-2ex}
    \caption{\mdseries Log-scale running time per join pattern in \texttt{stats\_ceb\_join}.}
    \label{fig:stats_ceb_join_bar}
\end{figure*}

\begin{figure*}[htbp]
    \centering
    \includegraphics[width=\columnwidth]{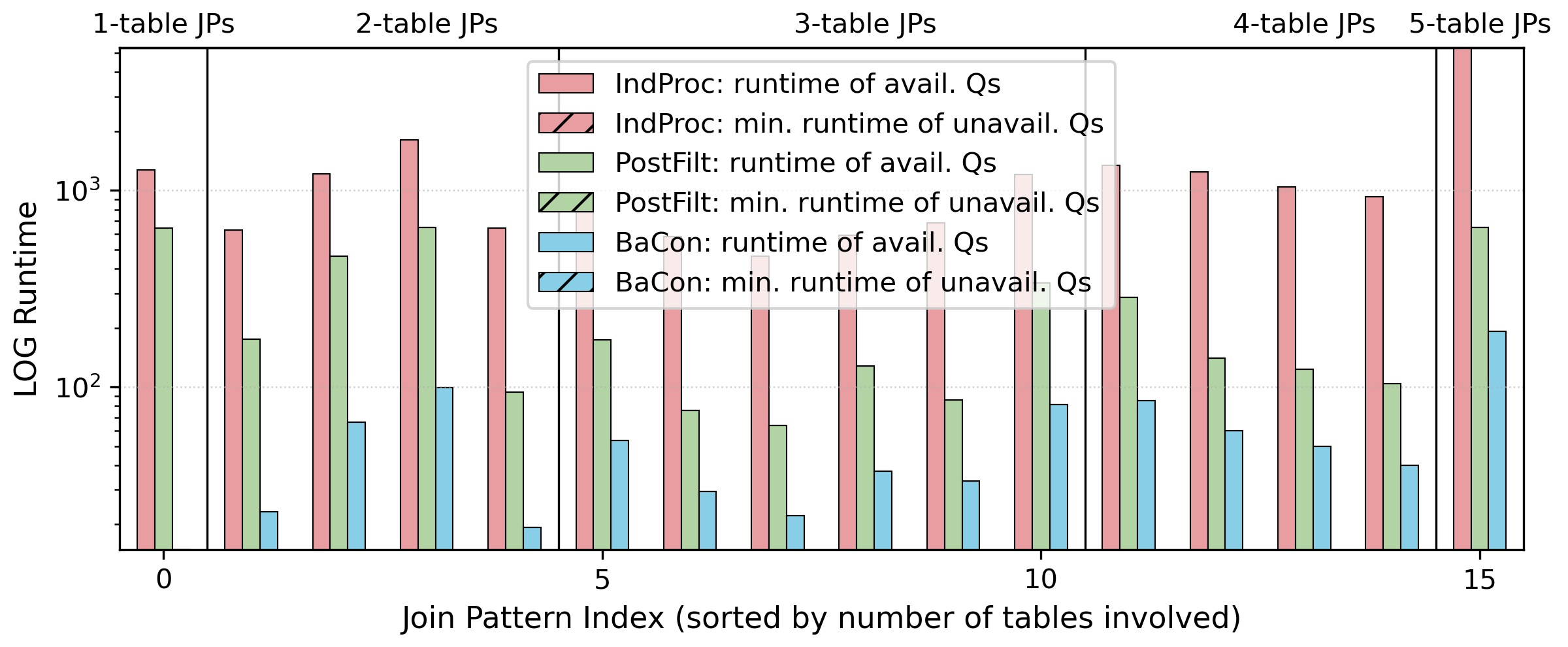}
    \vspace*{-2ex}
    \caption{\mdseries Log-scale running time per join pattern in \texttt{dsb\_grasp\_20k}.}
    \label{fig:dsb_grasp_20k}
\end{figure*}

\begin{figure*}[htbp]
    \centering
    \includegraphics[width=\columnwidth]{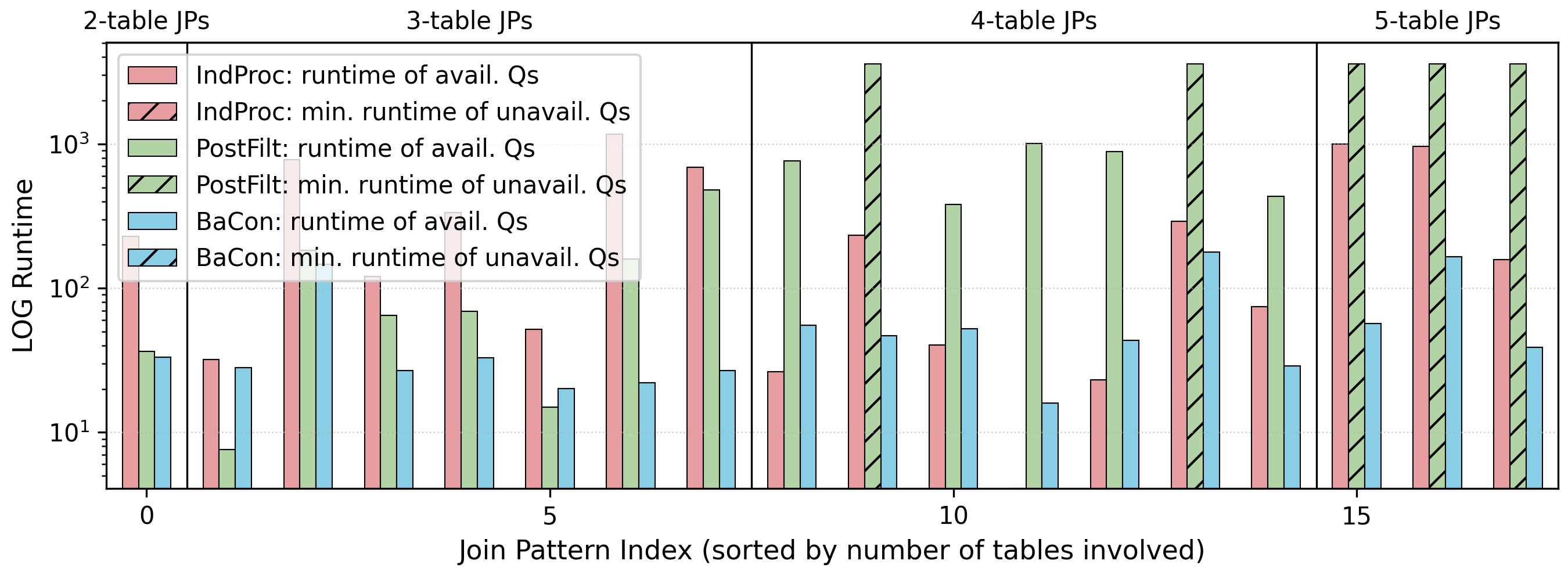}
    \vspace*{-2ex}
    \caption{\mdseries Log-scale running time per join pattern in \texttt{job\_light\_1k}.}
    \label{fig:job_light_1k_bar}
\end{figure*}

\begin{figure*}[htbp]
    \centering
    \includegraphics[width=\columnwidth]{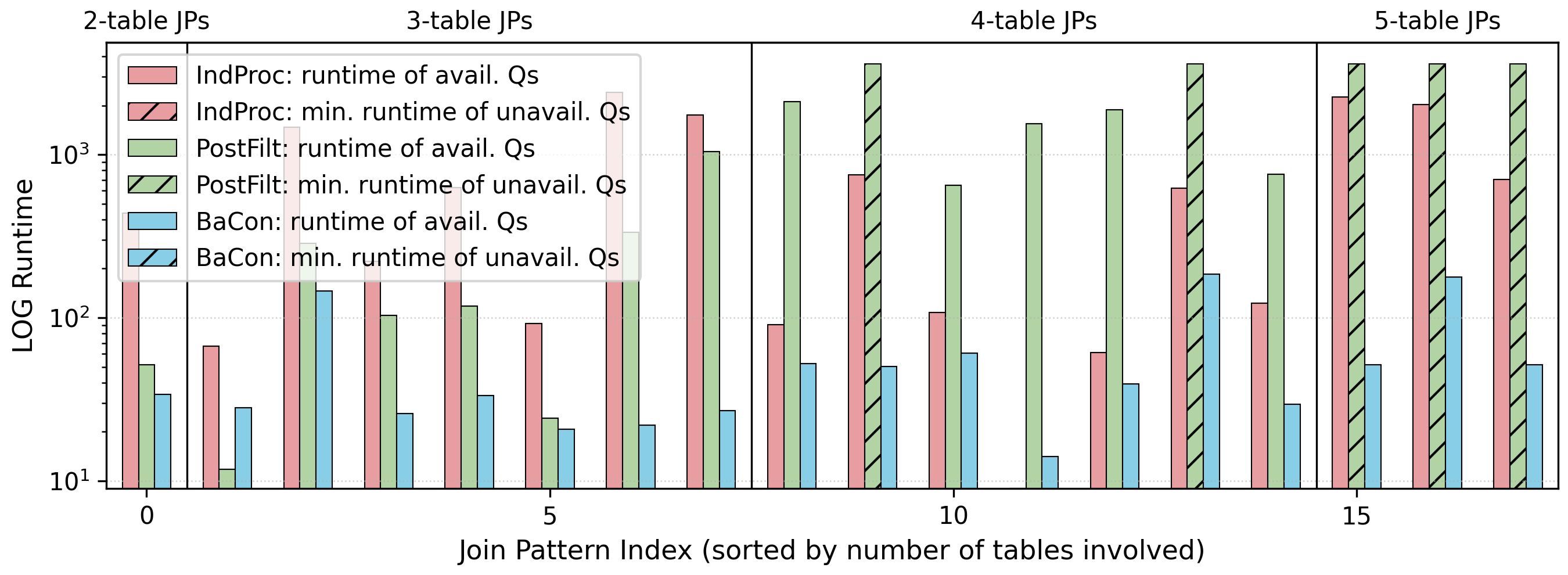}
    \vspace*{-2ex}
    \caption{\mdseries Log-scale running time per join pattern in \texttt{job\_light\_2k}.}
    \label{fig:job_light_2k_bar}
\end{figure*}

\begin{figure*}[htbp]
    \centering
    \includegraphics[width=\columnwidth]{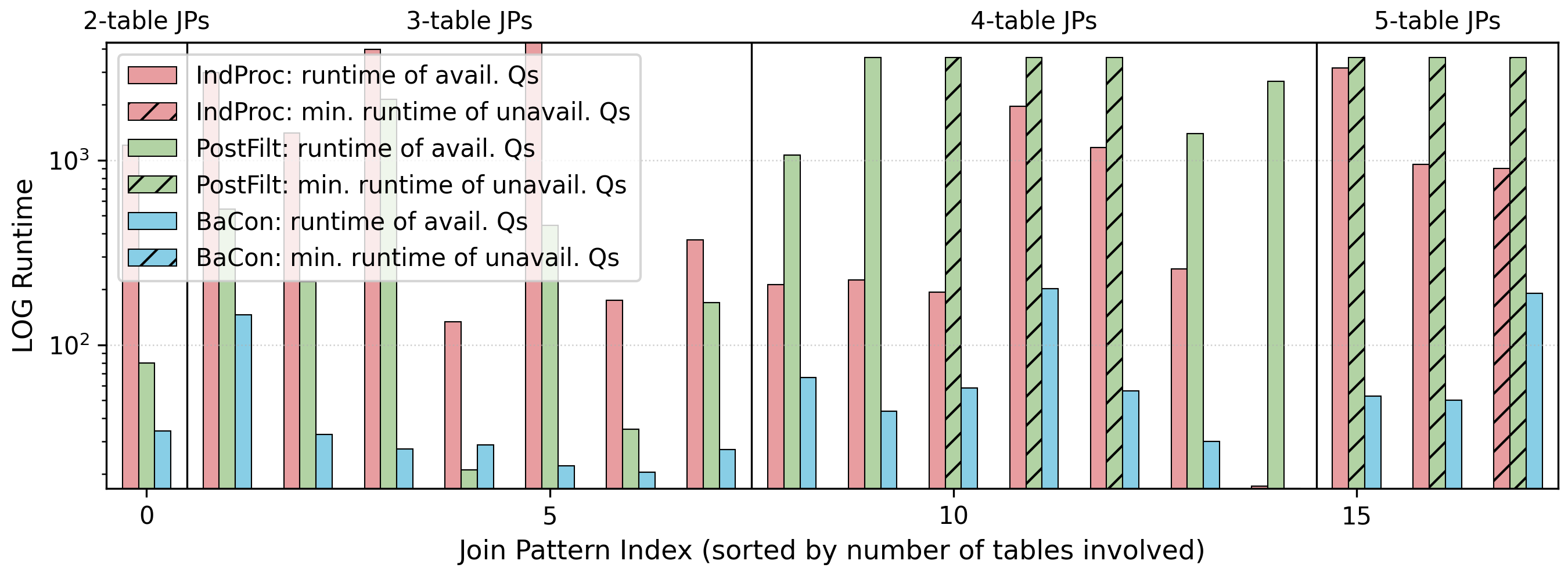}
    \vspace*{-2ex}
    \caption{\mdseries Log-scale running time per join pattern in \texttt{job\_light\_4k}.}
    \label{fig:job_light_4k_bar}
\end{figure*}
    \clearpage
    \section{More Details of Related Work}

\paragraph{Select-Join-Aggregate Query Processing.} Counting the join results (with selection predicate on the base tables) is a special class of select-join-aggregate queries (intuitively, applying an aggregate function on top of the select-join queries) with the output size as $1$. Processing such queries usually first push down the selection predicate to the base table, and ends up with processing join-count queries, which is the focus below.

The previous works have achieved two flavors of results: worst-case optimal and output-sensitive. Namely, worst-case optimal algorithms work only on pathological
instances with huge outputs, which are rare in practice. In contrast, output-sensitive algorithms express the runtime as a function of the input size and output size, which are more practically meaningful, especially for queries where the aggregation (such as count) may significantly reduce the output size. The classical
Yannakakis algorithm \cite{10.5555/1286831.1286840} can compute {\em free-connex} join-count queries in $O(N)$ time, where $N$ is the total number of tuples in the database. 
Very recently, \reva{\citet{10.1145/3725241} exploits} the hybrid strategy of Yannakakis algorithms, which can compute acyclic join-count queries in $O(N \cdot \mathsf{OUT}^{1-\frac{1}{\textsf{fnfhtw}}} + \textsf{OUT})$ time, where $\mathsf{OUT}$ is the size of the projection of the join results onto the counting columns, and $\textsf{fnfhtw}$ is the free-connex fractional hypertree width of the query. For general join-count queries, the state-of-the-art approach is to convert the query into a
free-connex one using the tree decomposition technique and the worst-case optimal join algorithm \cite{10.1145/3196959.3196990,ngo2012worstcaseoptimaljoinalgorithms,10.1145/2590989.2590991,veldhuizen2013leapfrogtriejoinworstcaseoptimal}, and then run the hybrid Yannakakis
algorithm on the tree decomposition. 

Our work is also inspired by some ideas in these works. 
The instantiations of worst-case optimal join algorithms~\cite{10.1145/3129246,10.14778/3407790.3407797,10.14778/3342263.3342643,10.1145/3589295} motivate \ours{} to decompose computation via child recursions and combine multiple coordinate-to-count maps during backtracking.
Cost-based optimizers in these systems, such as Free Join's~\cite{10.1145/3589295}, together with the AGM bound~\cite{atserias2017sizeboundsqueryplans,10.1145/2220357.2220363}, further inform our initial exploration of cost-based hybrid mechanism.
However, achieving more accurate cost estimates for \ours{} may conversely require more accurate (pessimistic) cardinality estimation at a higher cost, such as \reva{LpBound}~\cite{DBLP:journals/pacmmod/ZhangMKOS25}, and/or specialized data storages and execution engines, such as Free Join's COLT~\cite{10.1145/3589295} and its integration to DuckDB~\cite{Raasveldt2022DuckDBA}.

\paragraph{Aggregation Push-down}
Early works explore aggregation push-down in query plans to reduce intermediate result sizes.
Eager aggregation and lazy aggregation~\cite{10.5555/645921.673154} introduce rules for early or deferred partial aggregation based on algebraic properties, while\reva{~\citet{10.5555/645920.672834}} integrate \sql{GROUP BY} into cost-based optimization;
both focus on intra-query optimization via query rewriting. 
More recently, GuAo~\cite{10.14778/3718057.3718068} identifies aggregate queries that can be evaluated without materializing any joins and introduces a corresponding physical operator for SparkSQL~\cite{10.1145/2723372.2742797}.
GuAo targets single-query execution and assumes selection push-down at table scans.
For the case without \sql{GROUP BY}, its propagation of frequencies grouped by join attributes resembles \ours{}'s bucketization when reduced to a single-query join pattern.
Unlike \ours{}, however, GuAo merges frequencies via standard binary joins in SparkSQL rather than supporting multi-way combination during count-map merging.

\paragraph{Factorizations} 
Factorized databases (FDBs) propose the factorized representations of query results~\cite{10.14778/2350229.2350242} to eliminate tuple-level redundancy and boost the relational processing performance~\cite{DBLP:conf/icdt/OlteanuZ12}, supporting aggregations beyond \sql{COUNT}~\cite{10.14778/2556549.2556579}.
These works focus on intra-query redundancy in query output and are different but complementary to our approach.
Moreover, they assume storage layout different from traditional relational databases, like PostgreSQL. 


\paragraph{Optimized Batch Processing of Aggregate Queries}
\reva{LMFAO~\cite{10.1145/3299869.3324961} is a framework proposed} to efficiently compute a batch of group-by aggregates over shared joins without fully materializing intermediate results.
It decomposes queries into views, groups them according to a join tree, and evaluates each view group using a multi-output plan.
Similar to \postfilter{}, LMFAO supports selection predicates by encoding them as conditional expressions inside aggregate functions.
In contrast to \ours{}, LMFAO does not analyze or pre-process overlaps among cross-query predicates (e.g., via bucket construction), but instead focuses on sharing computation at the level of join tree traversal and attribute-ordered evaluation, reusing identical conditional expressions and partial products within a view group.
Moreover, LMFAO is implemented as a standalone engine, rather than being integrated into a traditional DBMS.



\paragraph{Multi-Query Optimization \& Shared Workload Optimization}
Multi-query optimization (MQO)~\cite{10.1145/42201.42203} identifies shared or similar subexpressions across queries and performs inter-query optimization by executing the common subexpressions jointly or by rewriting query plans with a boarder subquery.
Along this line, Zhou et al.~\cite{10.1145/1247480.1247540} propose a practical common subexpression manager integrated into Microsoft SQL Server and its cost-based query optimizer, demonstrating improvements on workloads of a few tens of queries. 
As the search space for identifying common subplans grow rapidly with the number of queries, Shared Workload Optimization (SWO)~\cite{10.14778/2732279.2732280} optimizes an entire workload -- often comprising hundreds or thousands of concurrent queries -- by identifying shared operators instead.
This typically requires specialized or substantially modified execution engines (e.g., shared work systems) to support shared physical operators and coordinated execution.
Both lines of work target more general queries and require accurate cost estimates, which motivates future extensions of \ours{} toward tighter, cost-based integration with DBMSs.

\paragraph{Continuous Query Processing}
Related works on continuous query processing~\cite{10.1145/564691.564698,10.5555/1182635.1164132} in stream systems explore shared computations by maintaining data summaries or predicate indices over evolving streams.
These approaches are designed for workloads with large numbers of continuous queries or filters (e.g., on the order of 100k), usually exceeding the workload sizes considered in our setting, and rely on specialized structures to efficiently support updates.
Nevertheless, they offer complementary insights that could inform strategies for handling workloads at different scales as future work.

\revc{
    \paragraph{Learned Cardinality Estimation} 
    As the main motivation of \ours{}, query-aware CE models~\cite{park2020quicksel,10.14778/3329772.3329780,10.14778/3476249.3476254,wu2021unified,kipf2018learned,10.14778/3476249.3476259,DBLP:conf/edbt/MullerWL23,10.1145/3514221.3517896,10.1145/3514221.3526179,10.14778/3626292.3626302} learns to predict cardinalities from a training set of query-count pairs.
    Training is typically performed either in a single-shot manner, where a model is (re-)trained from scratch on a collected batch of queries~\cite{10.14778/3329772.3329780, 10.1145/3514221.3526179,li2021cardinality,10.14778/3476249.3476254,10.1145/3514221.3517896}, or in multiple rounds, where the model is incrementally updated as new query-count pairs occur (e.g., via fine-tuning)~\cite{10.14778/3626292.3626302,10.1145/3654932,park2020quicksel}.
    The required input information varies across approaches.
    Pure data-driven CE models~\cite{10.14778/3368289.3368294,10.14778/3421424.3421432} learn data distributions directly from the tables and do not require query-count pairs for training, although query workloads are still used for evaluation.
    Some hybrid models~\cite{kipf2018learned,10.14778/3626292.3626302} additionally require data statistics, such as samples or histograms.
    Beyond model design, recent work has focused on model robustness~\cite{10.1145/3588713,10.14778/3725688.3725708}, training efficiency~\cite{jgmp}, and adaptivity to updates and drifts~\cite{10.1145/3514221.3526179,10.1145/3639293,10.14778/3583140.3583164,10.1145/3654932,10.14778/3626292.3626302}, which further increases the demand for diverse query workloads.
}
\end{appendix}

\end{document}